\documentclass[twocolumn,tightenlines,showpacs,prd,floatfix,preprintnumbers,amsmath,amssymb,nofootinbib,superscriptaddress]{revtex4-1}

\usepackage{bm}
\usepackage{graphicx}
\usepackage{float}
\usepackage{amsmath}
\usepackage{acronym}
\usepackage{multirow}
\usepackage{amsfonts}
\usepackage{subfigure}
\usepackage[usenames,dvipsnames]{xcolor}

\newcommand{\be}{\begin{equation}}
\newcommand{\ee}{\end{equation}}
\newcommand{\ba}{\begin{array}}
\newcommand{\ea}{\end{array}}

\begin{document}
\title{Comparison of Gravitational Wave Detector Network Sky Localization Approximations}

\author{K. Grover}
\affiliation{School of Physics and Astronomy, University of Birmingham, Edgbaston, Birmingham B15 2TT, UK}
\author{S. Fairhurst}
\affiliation{School of Physics and Astronomy, Cardiff University, Cardiff, CF24 3AA, United Kingdom  }
\author{B. F. Farr}
\affiliation{Northwestern University, Evanston, IL 60208, U.S.A. }
\affiliation{School of Physics and Astronomy, University of Birmingham, Edgbaston, Birmingham B15 2TT, UK}
\author{I.~Mandel}
\affiliation{School of Physics and Astronomy, University of Birmingham, Edgbaston, Birmingham B15 2TT, UK}
\author{C. Rodriguez}
\affiliation{Northwestern University, Evanston, IL 60208, U.S.A. }
\author{T. Sidery}
\affiliation{School of Physics and Astronomy, University of Birmingham, Edgbaston, Birmingham B15 2TT, UK}
\author{A.~Vecchio}
\affiliation{School of Physics and Astronomy, University of Birmingham, Edgbaston, Birmingham B15 2TT, UK}

\date\today

\begin{abstract} 
Gravitational waves emitted during compact binary coalescences are a promising source for gravitational-wave detector networks.  The accuracy with which the location of the source on the sky can be inferred from gravitational wave data is a limiting factor for several potential scientific goals of gravitational-wave astronomy, including multi-messenger observations. Various methods have been used to estimate the ability of a proposed network to localize sources. Here we compare two techniques for predicting the uncertainty of sky localization -- timing triangulation and the Fisher information matrix approximations -- with Bayesian inference on the full, coherent data set. 

We find that timing triangulation alone tends to over-estimate the uncertainty in sky localization by a median factor of $4$ for a set of signals from non-spinning compact object binaries ranging up to a total mass of $20 M_\odot$, and the over-estimation increases with the mass of the system.  We find that average predictions can be brought to better agreement by the inclusion of phase consistency information in timing-triangulation techniques.  However, even after corrections, these techniques can yield significantly different results to the full analysis on specific mock signals.  Thus, while the approximate techniques may be useful in providing rapid, large scale estimates of network localization capability, the fully coherent Bayesian analysis gives more robust results for individual signals, particularly in the presence of detector noise.

\end{abstract}

\maketitle
\section{Introduction}

Neutron star (NS) and black hole (BH) binaries will be an important source of gravitational waves (GWs) detectable by advanced ground-based laser interferometers that are expected to come online in 2015, the two LIGO detectors in the US and the Virgo detector in Italy \cite{AdvLIGO, AdvVirgo}. These instruments are sensitive to the final minutes to seconds of a binary coalescence and the detection rate is anticipated to be between 0.4 yr$^{-1}$ and 400 yr$^{-1}$ at final design sensitivity ~\cite{LVC2010}. This new class of observations is anticipated to provide us with new insights into the formation and evolution of this class of relativistic objects and their environments \cite[e.g.][]{Kalogera:2007, MandelOShaughnessy:2010, Li:2012}.

An important element in any astronomical observation is the ability to locate a source in the sky. Compact binaries that are detected via GWs provide a radically new sample of compact object binaries in a highly relativistic regime, which is otherwise difficult to identify. If some of the binding energy is released in the electromagnetic band as a prompt burst of radiation during the merger or as an afterglow, as is expected in the case of NS-NS and NS-BH coalescences, signatures in the gamma-ray, X-ray, optical, and radio spectra may be identified by telescopes, and will provide a multi-wavelength view of these phenomena \cite[e.g.][]{metzger2012}.  If, on the other hand, the merger is electro-magnetically silent, as expected for a binary black hole merger in vacuum, constraining the source location in the sky may provide unprecedented clues about the environments that harbor such exotic objects. 

Locating a GW source as precisely, and rapidly, as is technically possible is therefore an important element of GW observations. GW laser interferometers are not pointing telescopes, but the sky location can be reconstructed through the time of arrival of GW radiation at the different detector sites, i.e. {\it timing triangulation}, as well as the relative amplitude and phase of the GWs in different detectors, that carry additional information about the source geometry. During the final observational run (S6/VSR2-3) of the LIGO and Virgo instruments, in their so-called ``initial" configuration, a rapid sky localization algorithm based in part on timing triangulation was implemented and used to provide alerts on the time-scale of minutes to telescopes for follow-ups of detection candidates (none of which represented a confident GW detection) \cite{abadie12}. In parallel, a Bayesian inference analysis designed to provide accurate estimates of all the source parameters, including location in the sky, was deployed on a range of software and hardware injections, including the ``blind injection" \cite{LVC2013a}. Now that the instruments are undergoing upgrades \cite{AdvLIGO,AdvVirgo} a major effort is taking place to refine these sky localization techniques in preparation for observations at advanced sensitivity, and to estimate the pointing performance of the GW network in order to put in place an electro-magnetic (EM) follow-up observational strategy. 

When computing the typical size of the error-box in the sky, one needs to draw a careful distinction between actual parameter estimation on noisy data from a detector network, and predictions of the expected parameter-recovery accuracy in principle. Ultimately, one would like to know the probability that the source is enclosed within a given sky area in a real analysis. This information is provided by the posterior probability density function of the sky location given the data and all the assumptions of the model. For a multi-dimensional and possibly multi-modal parameter space, as is the case of models that describe GWs from coalescing binaries, this function can be efficiently obtained using a variety of stochastic sampling techniques.  Markov-chain Monte Carlo \cite{vanderSluys2008} and nested sampling \cite{veitch10} have both been applied to inference on GW data; in fact, a sophisticated software library specifically developed for this goal is now in place and actively developed \cite{LAL}.   Several assessments of the localization capability of the trans-continental network of GW instruments in different observational scenarios have recently been made using full, coherent parameter estimation techniques \cite[e.g.][]{veitch2012, Nissanke:2013, kasliwal13}.

However, the exploration of a large-dimensional parameter space can be computationally intensive and lengthy. Other methods have therefore been devised to estimate theoretically the uncertainty of the sky location recovery without going through the actual analysis process.   

One technique that has been used in countless studies is based on the evaluation of the Fisher information matrix (FIM), whose inverse provides an estimate of the covariance matrix on the parameter space \cite{cutler1994}. However, the FIM provides robust estimates only for a particularly a sufficiently high signal-to-noise ratio to validate the approximation of the likelihood as a Gaussian. It is now clear that it is of limited use to describe the actual performance of the network of ground-based laser interferometers in realistic conditions, see \cite{Vallisneri:2008} and most recently \cite{rodriguez13}.  

A different approach that has been considered specifically for predicting the sky localization error-box is based on using time delays between the arrival of GWs at different detectors in the network and theoretical estimates of time-of-arrival measurement uncertainty in individual detectors  \cite{fairhurst2009, wen2010}. A predictive method using solely timing accuracy is therefore based on the assumption of independent rather than coherent analyses of data streams from different detectors, and potentially misses out on additional information from a coherent analysis, such as the relative phase and amplitude of the GW strain between detectors, that could improve localization accuracy. This timing triangulation approach \cite{fairhurst2011} was used as the basis of the recent LIGO Scientific Collaboration and Virgo Collaboration document on prospects for localization of GW transients with advanced observatories \cite{LVC2013}.

The main question that we want to address in this paper is the following: how accurate are the predictions based solely on triangulation estimates with respect to what could be actually achieved in a real coherent analysis of a stretch of data in which a GW candidate has been identified? We answer this question by considering a set of synthetic gravitational-wave signals from coalescing compact binaries added to Gaussian and stationary noise representative of the sensitivity achieved in the last LIGO -- Virgo science run.  We then examine and compare the predicted sky localization uncertainty estimates obtained with different techniques for this common set of simulated sources. Our study measures the accuracy of estimates from timing triangulation and Fisher information matrix analyses relative to the optimal coherent Bayesian inference, which defines what is achievable in practice with a given data set and model.

The paper is organized as follows. Section~\ref{sec:methods} outlines the methods used to calculate the sky areas. Section III applies these methods to a set of simulated GW signals. We  compare the three methods in Section IV, and conclude with a discussion in Section V.

\section{Methods}
\label{sec:methods}

The data, $d_j(t)$, from each GW detector $j = 1,\dots, N$, where $N$ is the number of instruments, is a sum of the noise $n_j(t)$ and any signal $h(t;{\bm\theta})$,
\begin{equation}
	\label{eqn:data}
	d_j(t) = n_j(t) + h_j(t;{\bm\theta})\;,
\end{equation}
where ${\bm \theta}$ is a vector that describes the set of unknown parameters that characterize the emitting source. The waveforms for coalescing binary systems $h({\bm\theta})$ used in this study are discussed in Section \ref{sec:det_inj}. The noise in each detector is assumed to be a zero-mean, stationary, Gaussian process characterized by a one-sided noise spectral density $S_{n_j}(|f|)$. The signal, $h(t;{\bm\theta})$, from compact binary coalescences (CBCs) with non-spinning components can be described with a nine-dimensional parameter vector ${\bm \theta}$: the two component masses $m_1$ and $m_2$ (\,or, alternatively, the symmetric mass ratio $\eta = {m_1m_2}/{\left( m_1+m_2\right)^2}$ and the chirp mass $\mathcal{M}_c = \eta^{3/5}\left(m_1+m_2\right)$\,), the distance to the source $D$, the source location in the sky -- right ascension $\alpha$ and declination $\delta$, the orientation of the binary -- polarization $\psi$ and inclination $\iota$, and the reference phase $\phi_0$ and time $t_0$. 
It is useful to define the noise-weighted inner product between two functions $a$ and $b$ as
\be
	\left \langle a|b \right \rangle = 2 \int_0^\infty df \frac{\tilde a^*(f) \tilde b(f) + \tilde a(f) \tilde b^*(f)}{S_n(f)}\, ,
	\label{eqn:inner_prod}
\ee
where the integral is formally over all positive frequencies, but in practice is restricted, through $S_n(f)$, to the finite bandwidth over which the instrument is sensitive.

\subsection{Bayesian Parameter Estimation}
\label{sec: bayesian method}

The parameter-estimation pipeline of the LIGO -- Virgo collaboration uses a Bayesian framework to coherently analyze data from all detectors in the network \cite{LVC2013a}. Bayes' law gives the posterior probability density function $p\left( {\bm \theta} \mid d\right)$ in terms of the likelihood $p\left(  d \mid {\bm \theta} \right)$ and prior $ p\left( {\bm \theta}\right)$:
\begin{equation}
	\label{eqn:bayes law}
	p\left( {\bm \theta} \mid d\right) = \frac{ p\left( {\bm \theta}\right)  p\left(  d \mid {\bm \theta} \right)}{p\left( d \right)} \, ,
\end{equation}
where the denominator $p(d)$ is the evidence, which we treat as a normalization factor.  In stationary, Gaussian noise, the likelihood is

\begin{equation}
	\label{likelihood}
	p\left(d \mid {\bm \theta} \right) = \exp\left[-\frac{1}{2} \left \langle d -h({\bm \theta}) \mid d -h({\bm \theta}) \right\rangle\right]\, ,
\end{equation}
where $\left\langle ...\mid...\right\rangle$ denotes the inner product defined in Eq.~(\ref{eqn:inner_prod}). This is simply extended to the detector network analysis where the likelihood becomes a product of the likelihoods in individual detectors. 

The Bayesian pipeline maps out the posterior probability density function of the signal parameters given the data and model, and thus directly provides the statistical measurement uncertainty on parameter estimates. The sampling algorithm used in this paper is the LAL \cite{LAL} implementation of MCMC \cite{vanderSluys2008}, known as \texttt{LALinference\_mcmc}. It explores the parameter space using a random walk based on Metropolis-Hastings sampling methods \cite{gilks1996}. The performance of this pipeline has been demonstrated on a number of simulated events \cite{LVC2013a}, the sky localization performance investigated (on the same set of simulated sources as here) in \cite{sidery13} and the sky localization, among other parameters, investigated for  binary neutron star sources at high signal-to-noise ratios (SNRs) in \cite{rodriguez13a}. The code returns samples from the posterior which are binned using a $k$D-tree algorithm \cite{sideryKD} to compute the relevant two-dimensional probability intervals on the sky marginalized over all other parameters. For the comparisons reported below, we considered the smallest area which encloses $50\%$ of the total posterior probability $A_{\mathrm{Bayes}}$, also known as the $50\%$ credible interval. 

\subsection{Timing Triangulation Approximation}
\label{sec: TT method}

The Bayesian methods described above are computationally expensive; therefore, approximate techniques capable of analyzing a large set of synthetic GW sources and predicting their localization accuracies are desirable. In particular, a timing triangulation (TT) approximation, outlined in \cite{fairhurst2009} and \cite{fairhurst2011}, has been used to predict the sky localization ability of the upcoming advanced LIGO-Virgo network \cite{LVC2013}. TT assumes that the bulk of information that allows for source localization on the sky is contained in the relative time delays of the arrival of the GW train at different detectors in the network, and so uses estimates of time-of-arrival uncertainties in individual detectors to predict the overall localization uncertainty \cite{fairhurst2009}. The timing uncertainty in each detector $\sigma_t$ is estimated to be:
\begin{equation}
	\label{eqn: sigma t}
	\sigma_t= \frac{1}{2\pi \rho \sigma_f}\, ,
\end{equation}
where $\rho$ is the expectation value of the SNR in the detector $\rho^2=\left \langle h \mid h \right \rangle$ and the effective bandwidth of the signal in the detector $\sigma_f$ is
\begin{equation}
	\label{eqn: sigma f}
	\sigma_f^2=\overline{f^2} - \left(\;\overline{f}\;\right) ^2 \, ,
\end{equation}
where 
\begin{equation}\label{eqn: fbar}
\overline{f^n} = \frac{1}{\rho^2} \left \langle f^n h\mid h\right\rangle\, .
\end{equation}

The network localization accuracy, described by the matrix $\mathbf{M}$ (whose inverse is the covariance matrix on the sky), is a function of the timing uncertainty $\sigma_{t_i}$ in each detector and the pairwise separation vectors of the detectors in the network $\mathbf{D}_{i,j} = \bm{d}_i-\bm{d}_j$, where $\bm{d}_i$ is the light travel time between detector $i$ and the geocenter:%
\begin{equation}
	\label{eq sky loc matrix}
	\mathbf{M}= \frac{1}{\sum_{k} \sigma_{t_k}^{-2}}\sum_{i,j} \frac{\mathbf{D}_{i,\,j}\mathbf{D}_{i,\,j}^T }{2\sigma_{t_i}^2\sigma_{t_j}^2}\;,
\end{equation}
where $i,j,k$ label the detectors. 

We use the sky area as defined for the three-detector network in equation (33) of \cite{fairhurst2009} to calculate the $P=50\%$ credible interval, 
\begin{equation}
	\label{triang}
	A_{\mathrm{TT}}(P) = -2 \pi \sigma_x \sigma_y \frac{ \ln(1 - P)}{\cos\gamma}\;,
\end{equation}
where $\sigma_x$ and $\sigma_y$ are the inverse square roots of the two non-zero eigenvalues of $\mathbf{M}$ and $\gamma$ is the angle between the direction to the source and the normal to the plane containing the three detectors. 

\subsection{Fisher Information Matrix Approximation}
\label{sec: fisher method}

The calculation of the timing uncertainty, Eq.~(\ref{eqn: sigma t}), in the TT method described above is in essence a two-dimensional (time and phase) application of the Fisher information Matrix (FIM) technique. The FIM is an approximation with a long history of use in GW data analysis \cite[e.g.,][]{cutler1994}. In the high SNR limit, the likelihood over the desired credible region can be approximated as a Gaussian centered on the true parameters:
\begin{equation}
	\label{eqn: fisher dist}
	p(d | {\bm \theta}) \propto e^{-\frac{1}{2} \left \langle h_a  | h_b \right\rangle \Delta \theta^a \Delta \theta^b }\, ,
\end{equation}
where  $\Delta \theta _a = \theta_a - \theta^{\textrm{true}}_a $,  $h_a =  \frac{\partial h}{\partial \theta^a}$and repeated indices are summed over. 

In this case the Fisher information matrix is
\begin{equation}
	\label{eqn: fisher}
	\Gamma_{ab} \equiv \left\langle \frac{\partial h}{\partial \theta^a} \middle | \frac{\partial h}{\partial \theta^b} \right\rangle\;,
\end{equation}
where $h$ is the gravitational-wave signal and the inner product is defined in Eq.~(\ref{eqn:inner_prod}). The FIM for a multi-detector network is a sum of individual-detector FIMs. The FIM approximation has been found to be useful for large-scale investigations into sky localization of detector networks, although individual sky areas estimated via FIM can be orders of magnitude different from full coherent analysis. \cite{rodriguez13}.  Assuming flat priors the covariance matrix is then,
\begin{equation}
	\label{eqn:covar}
	\left \langle  \Delta \theta^a \Delta \theta^b \right \rangle = (\Gamma^{-1})^{ab} \equiv \Sigma^{ab}.
\end{equation}

We use the solid angle sky area as defined in equation (43) of \cite{barack2004} to calculate the $P=50\%$ credible interval,
\begin{equation}
	\label{fisher area}
	A_{\mathrm{F}}(P)=-2\pi \cos\left(\delta^{\textrm{true}}\right){\left[\ln{\left( 1- P \right)} \right]}\sqrt{ \Sigma^{\alpha \alpha} \Sigma^{\delta \delta} - \left( \Sigma^{\alpha \delta}  \right)^2   }\;,
\end{equation}
where $\alpha$ and  $\delta$ label the right ascension and declination respectively, $\delta^{\textrm{true}}$ is the declination of the source and here indices are not summed over. 
Below, we describe two different FIM analyses.  In the first, the full 9-dimensional parameter space described at the beginning of section \ref{sec:methods} is considered.  The $9$-dimensional FIM and its inverse, the $9$-dimensional covariance matrix, are used to estimate the area $A_{\mathrm{F9}}$ via Eq.~(\ref{fisher area}).  

As an interesting comparison we also include FIM results based on a reduced four-dimensional analysis incorporating only the two sky location parameters of the source (the right ascension and declination) as well as the reference phase and time of merger, with the remaining parameters held fixed, i.e., assumed to be known perfectly. Such an analysis artificially restricts the parameter space and assumes perfect knowledge of parameters which are correlated with sky location.  The $50\%$ credible level sky localization areas computed from the four-dimensional FIM approximation are reported as $A_{\mathrm{F4}}$.

\section{Detector Network and Injections}
\label{sec:det_inj}

\begin{figure}[h]
	\includegraphics[scale=0.45]{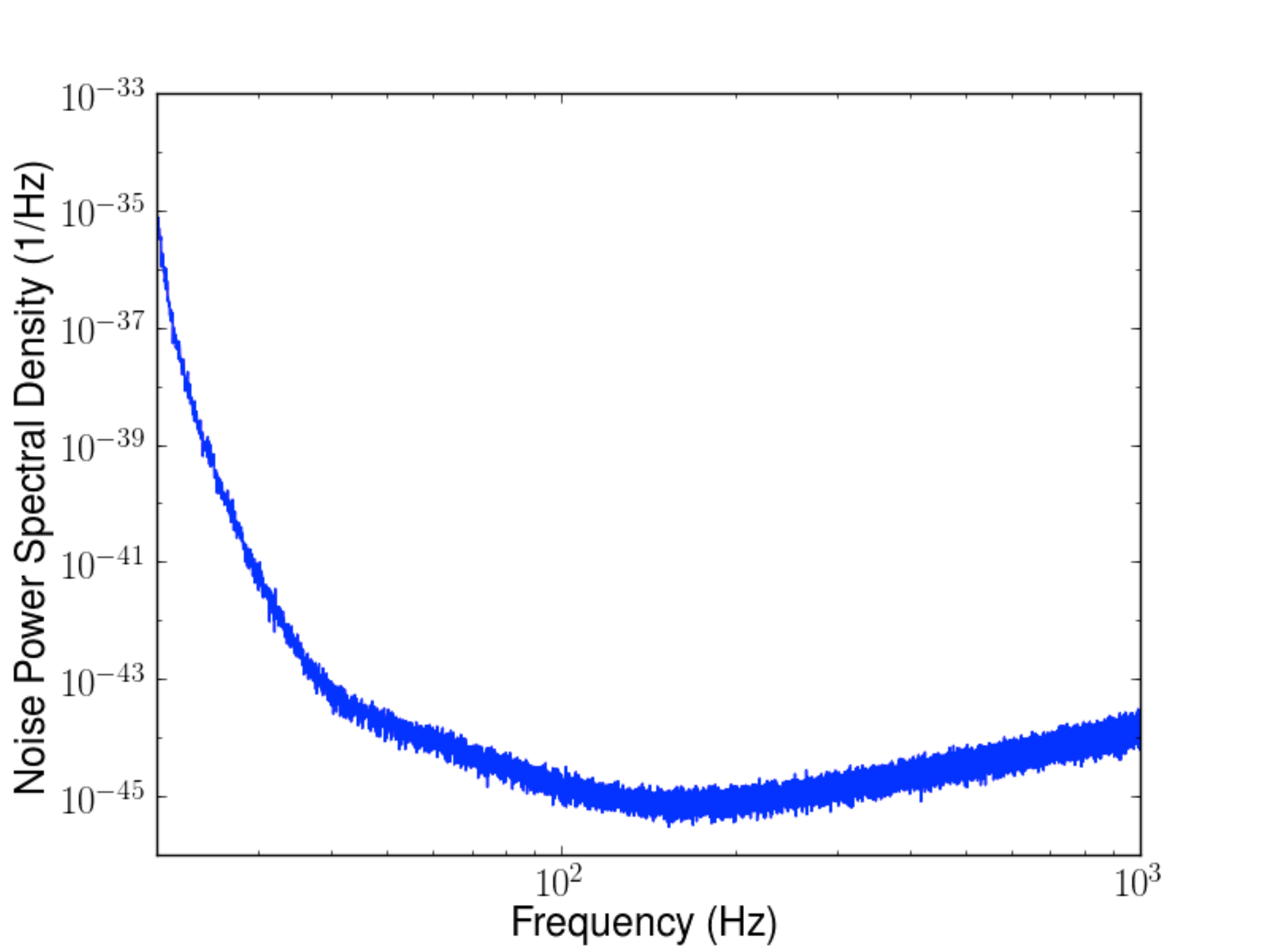}
	\caption{The one-sided noise power spectral density $S_n(|f|)$ as estimated from the mock data. The same noise spectral density, which is representative of the noise during the LIGO S6 operational period, was used to generate noise realizations in each of the three instruments (LIGO Hanford, LIGO Livingston and Virgo) that make up the detector network used in the study.}
	\label{fig:psd}
\end{figure}

To compare the methods described in the previous section we consider a set of mock observations made with the LIGO -- Virgo detector network. The network is composed of three detectors located at the LIGO Hanford (Washington state, USA) LIGO Livingston (Louisiana, USA) and Virgo (near Pisa, Italy) sites. We model the noise in each instrument as zero-mean, Gaussian and stationary with the same one-sided power spectral density $S_{n_j} = S_n(|f|)$ for each instrument, $j=1,2,3$. We use the noise power spectral density typical of the sensitivity achieved during the last science run (LIGO S6)  \cite{ligo12}; Figure~\ref{fig:psd} shows the noise power spectral density.

The signal model $h(t;{\bm \theta})$  describes the inspiral phase of a low-mass compact-object binary in circular orbit. The waveform is generated in the time domain and is modeled using the restricted post-Newtonian (pN) approximation, with amplitude at the leading Newtonian order and phase at the 3.5 pN order. The compact objects are assumed to be non-spinning and the specific waveform approximant used for injection and Bayesian recovery is TaylorT4 \cite{buonanno09}. The waveform used in the FIM analysis is the frequency-domain TaylorF2 approximant, which uses the restricted post-Newtonian approximation with amplitude at the leading Newtonian order and phase at the 2pN order \cite{buonanno09}.  Given the noise spectral density adopted for this study, the waveforms are generated from a starting frequency of $40$ Hz and are terminated at the usual condition for the these approximants, when the waveform reaches the innermost stable circular orbit frequency, $f_{\mathrm{ISCO}}=1/[6^{3/2}\pi (m_1+m_2)]$ . In all cases, we neglect the merger and ringdown parts of the waveform even though, for the higher mass systems, they would contribute to the signal power and to the source localization.

\begin{figure}[h]
                \includegraphics[width=0.45\textwidth]{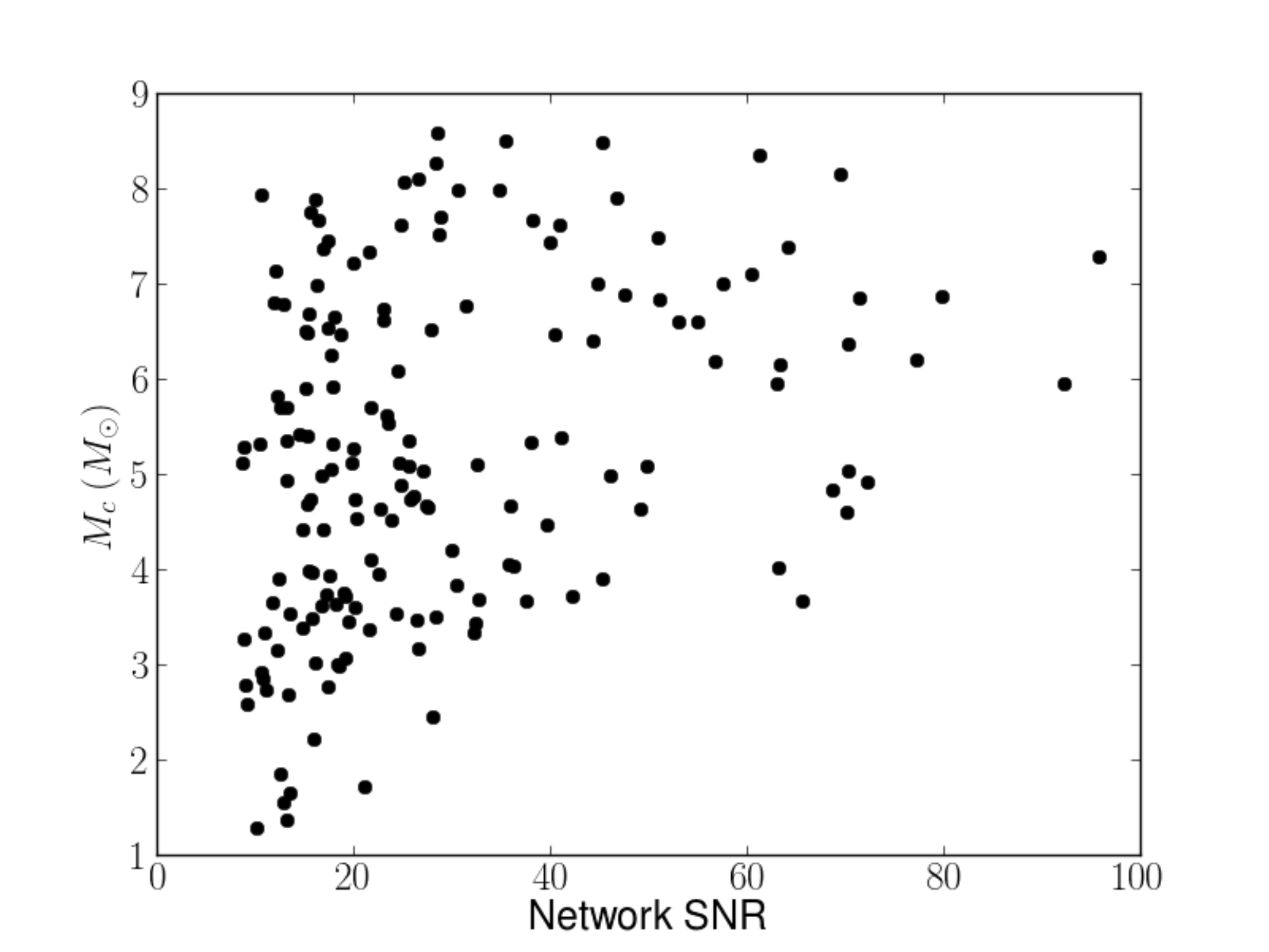}
                \caption{The chirp mass and coherent network signal-to-noise ratio of the set of sources being used to compare the sky localization methods.}
                \label{fig:chirp_injections}
\end{figure}

The set of signals added to the Gaussian noise were generated from parameters drawn randomly from the following population: the distance is distributed linearly in $\log(D)$ in the range $10-40$ Mpc, the location and orientation are isotropic on the respective two-spheres. The component masses $(m_1, m_2$) are drawn uniformly from $ 1\,M_\odot \le m_{1,\,2} \le15\,M_\odot $ with an additional cut on the total mass at $m_1 + m_2 \le 20\,M_\odot$. These injection distributions are also used as the priors in the Bayesian analysis. A total of 200 injections were generated. However, in this study we consider only 166 of the 200, based on the following two cuts.  We crudely emulate the detection pipeline by considering only signals with  $\rho \ge 5$ in at least two detectors.  We also consider only signals whose sky position was at least 5 degrees away from the plane defined by the detector network in order to use the simple TT approximation in Eq.~(\ref{triang}). 
The combination of distance and masses of the sources selected for the analysis leads to a distribution of chirp mass and total network SNR $\rho_{\mathrm{network}}= \sqrt{\sum_j{\rho_j^2} }$ that is shown in Figure~\ref{fig:chirp_injections}.

\section{Comparison of Methods}

\begin{figure*}[ht]
 	\begin{center}
	\subfigure[$\,$Full coherent bayesian]{
		\label{fig:hist_mcmc}
		\includegraphics[width=0.49\textwidth]{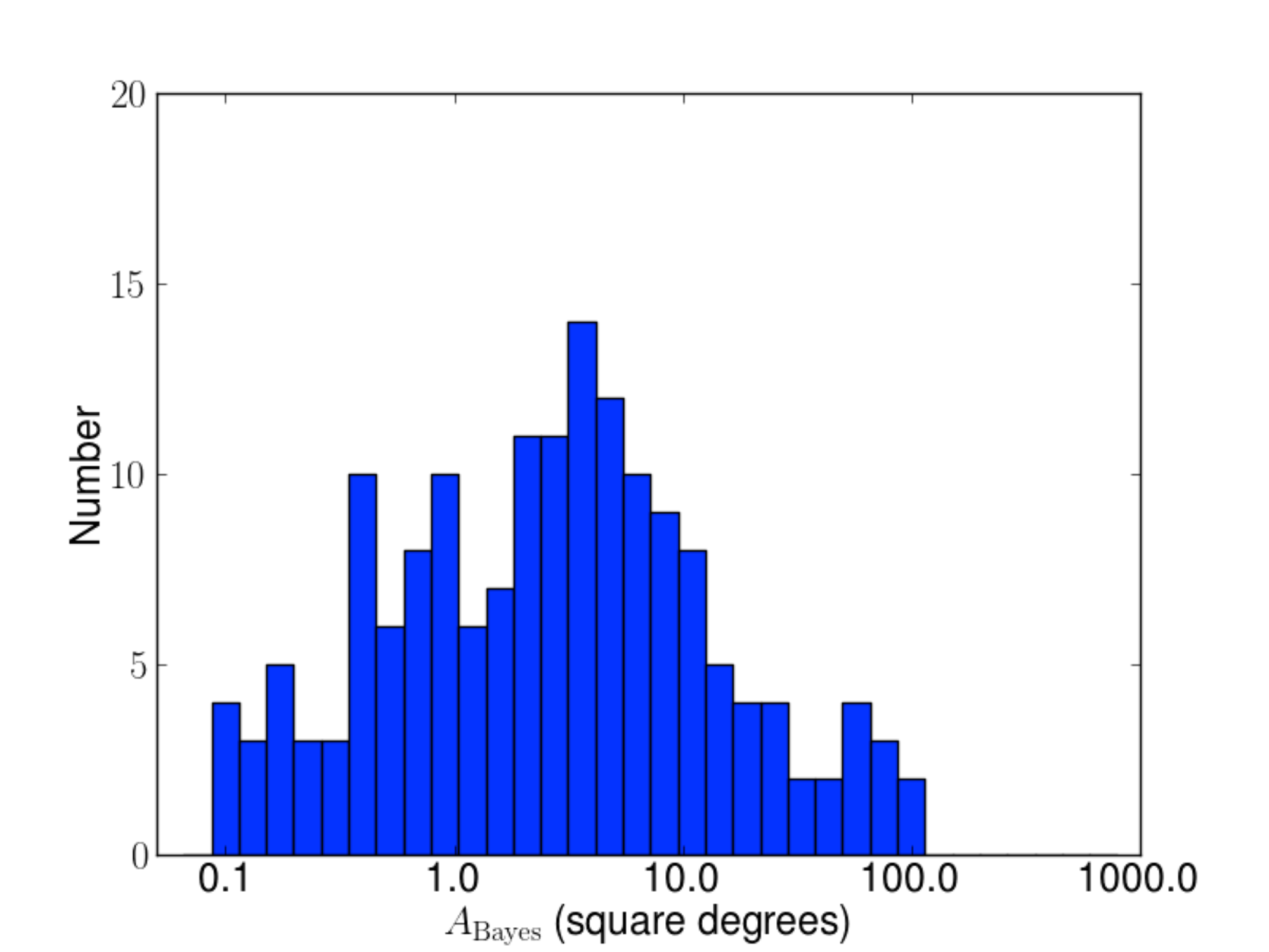}
        		}%
        \subfigure[$\,$Timing triangulation approximation]{
		\label{fig:hist_triang}
		\includegraphics[width=0.49\textwidth]{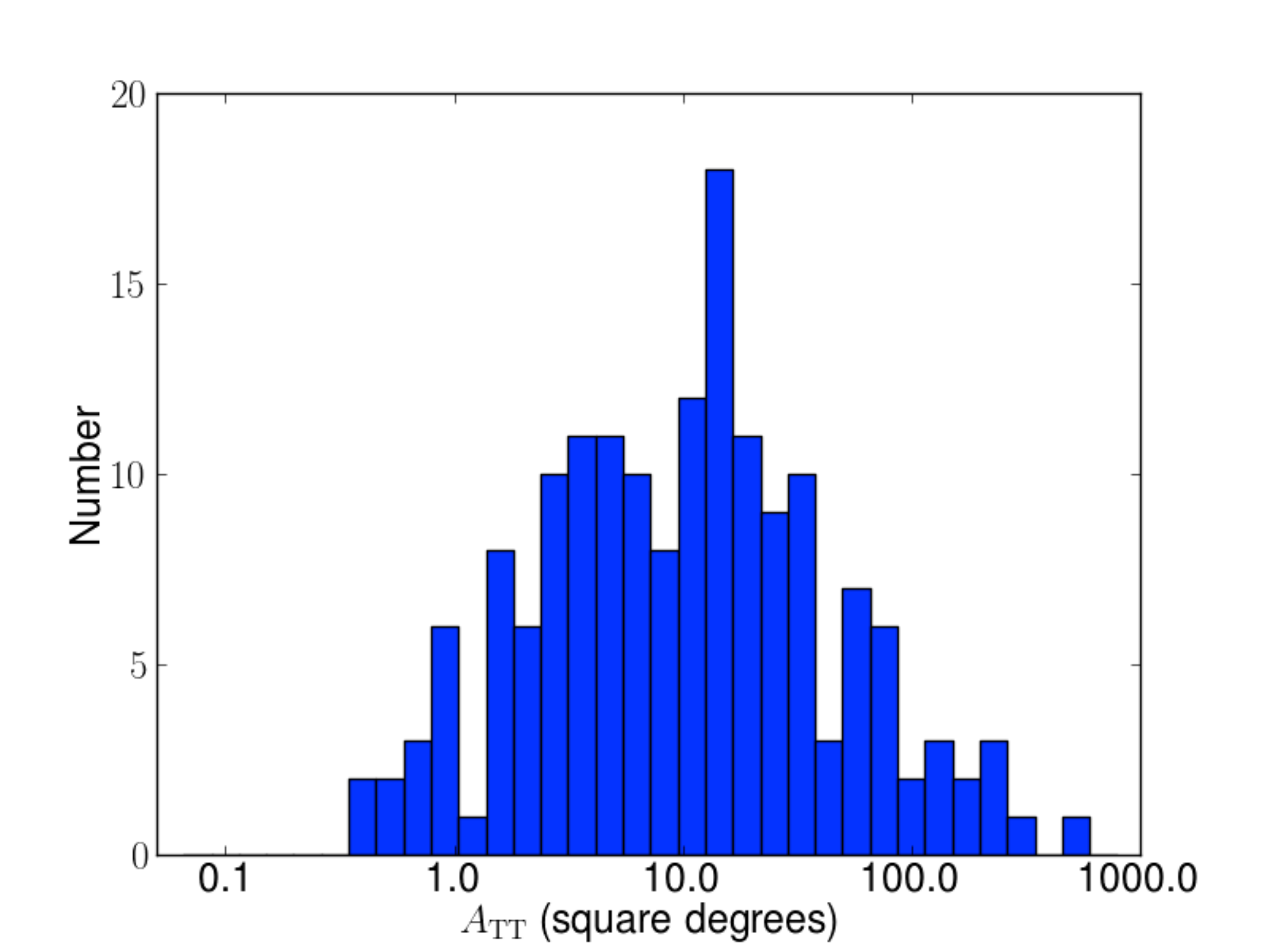}
		}\\		
	\subfigure[$\,$Fisher information matrix approximation 9D]{
		\label{fig:hist_fisher_9d}
		\includegraphics[width=0.49\textwidth]{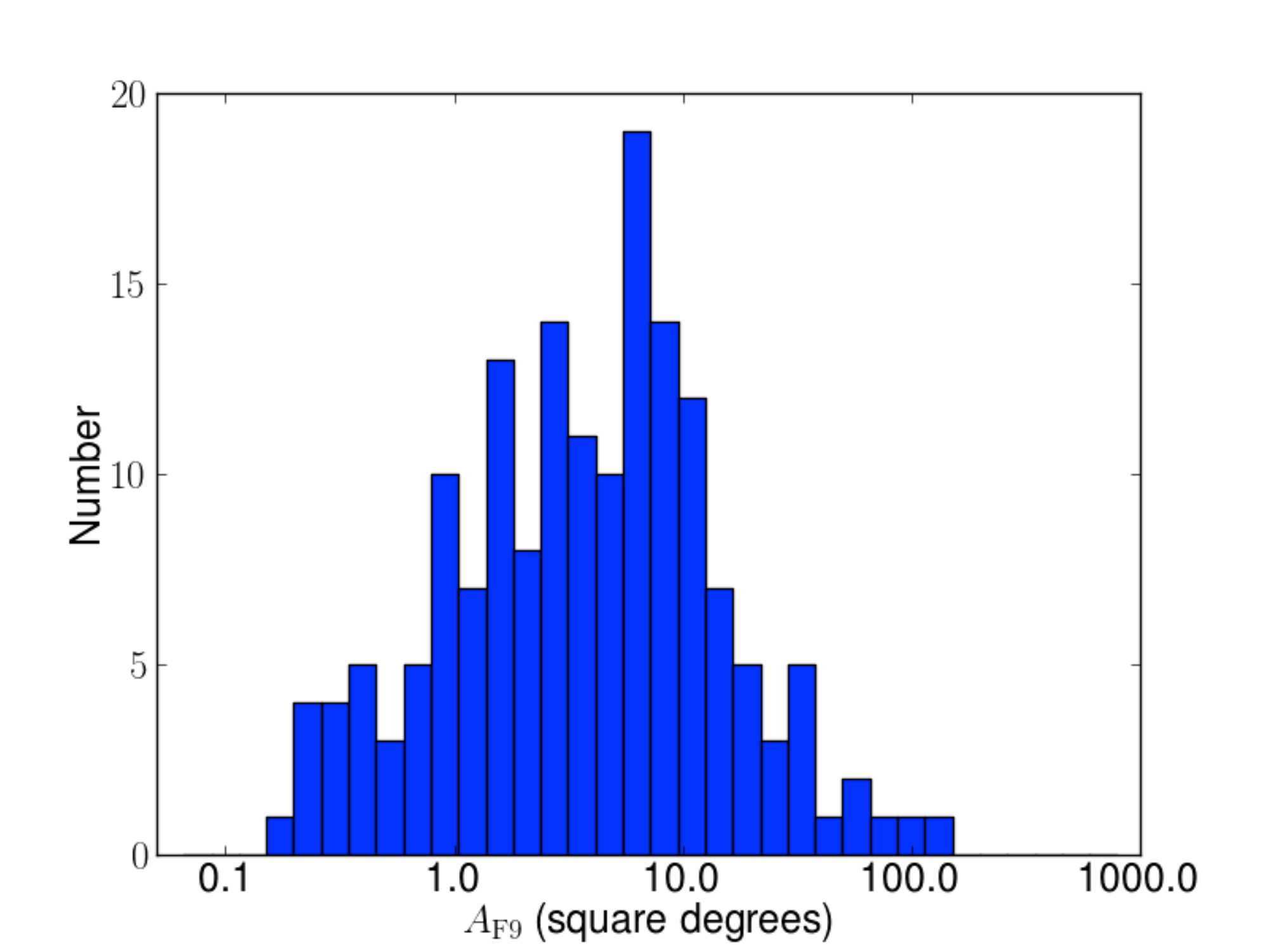}
        		}%
        \subfigure[$\,$Fisher information matrix approximation 4D]{
		\label{fig:hist_fisher_4d}
		\includegraphics[width=0.49\textwidth]{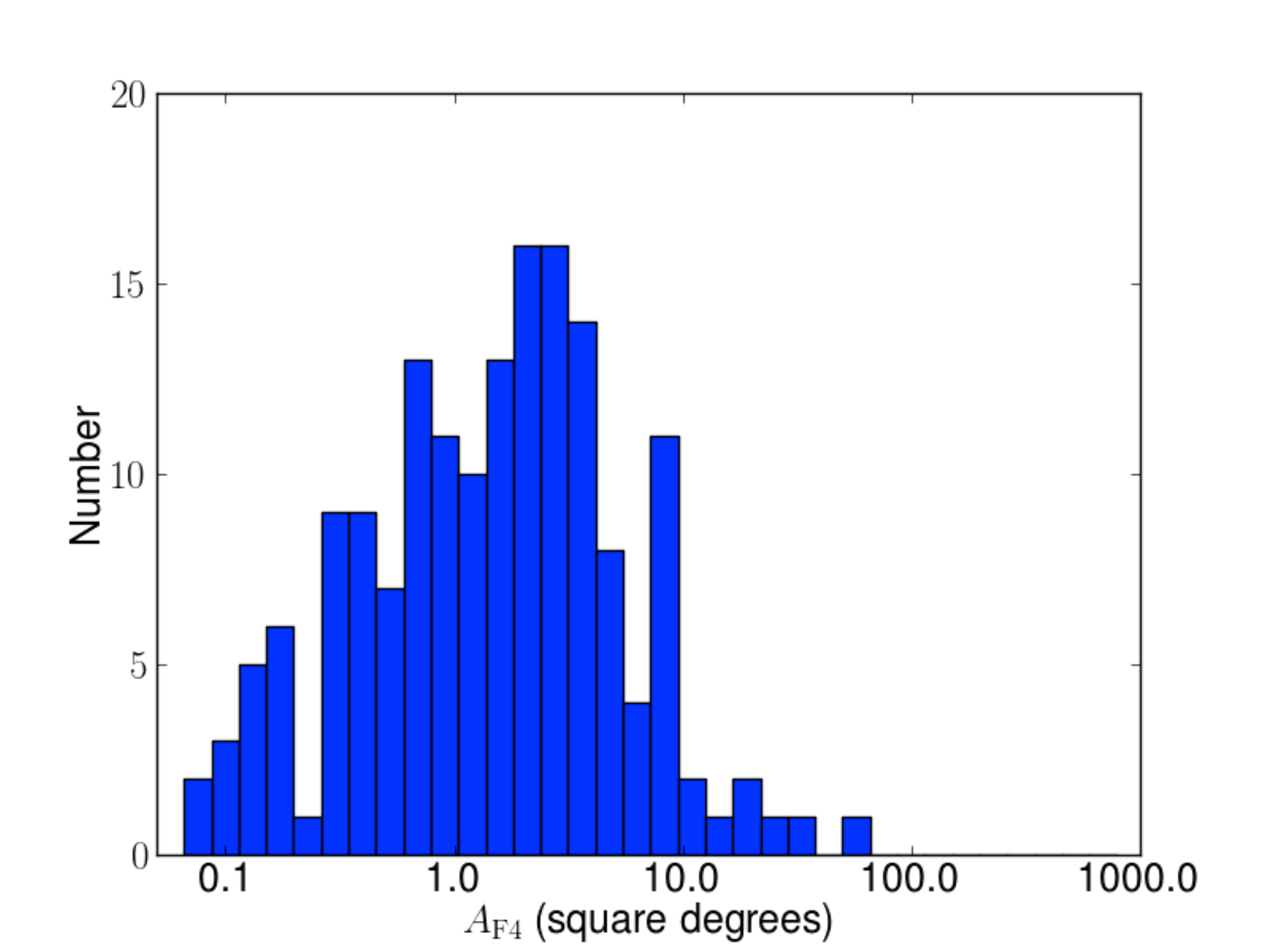}
        		}
	\end{center}
    	\caption{Histograms of the estimated $50\%$ credible-interval sky areas, in square degrees, calculated using the Bayesian coherent analysis $A_\mathrm{Bayes}$, timing triangulation $A_\mathrm{TT}$, 9-dimensional Fisher information matrix  $A_\mathrm{F9}$ and 4-dimensional Fisher information matrix, $A_\mathrm{F4}$.
	} 
	\label{fig:subfigures hists}
\end{figure*}

For each of the 166 signals described in the previous section, we computed the associated sky area at the 50\%  credible level (CL) using each of the four methods: coherent Bayesian analysis ($A_{\mathrm{Bayes}}$), timing triangulation ($A_{\mathrm{TT}}$), 9-dimensional Fisher information matrix ($A_{\mathrm{F9}}$) and 4-dimensional Fisher information matrix ($A_{\mathrm{F4}}$), as described in Section~\ref{sec:methods}. Figures \ref{fig:hist_fisher_4d}-\ref{fig:hist_mcmc} and Table~\ref{tab:summary} summarize the results. 

\begin{table}[h]
	\begin{center}
	\begin{tabular}{cccc}
		\hline
		\hline
		Method & 	Median & Mean & Standard Dev. \\
		 & (sq. deg.) &(sq. deg.) & (sq. deg.)\\
		\hline
		$A_{\mathrm{Bayes}}$  & 2.9 & 8.9 &17.1\\
		$A_{\mathrm{TT}}$ & 10.6 & 29.3 &59.3\\
		$A_{\mathrm{F9}}$& 4.0 & 8.7&16.0\\
		$A_{\mathrm{F4}}$& 1.6 & 3.2 &6.1\\
 		\hline
		\hline
	\end{tabular}
	\end{center}
	\caption{ Average $50\%$ credible-interval sky areas, in square degrees, computed using the four methods in Section \ref{sec:methods}. \label{tab:summary}} 
\end{table}

\subsection{Timing vs Coherent Bayesian}

\begin{figure*}[ht]
	\begin{center}
	\subfigure[$\,$Timing triangulation and full coherent Bayesian comparison ]{
		\label{fig: fairhurst /mcmc}
		\includegraphics[width=0.49\textwidth]{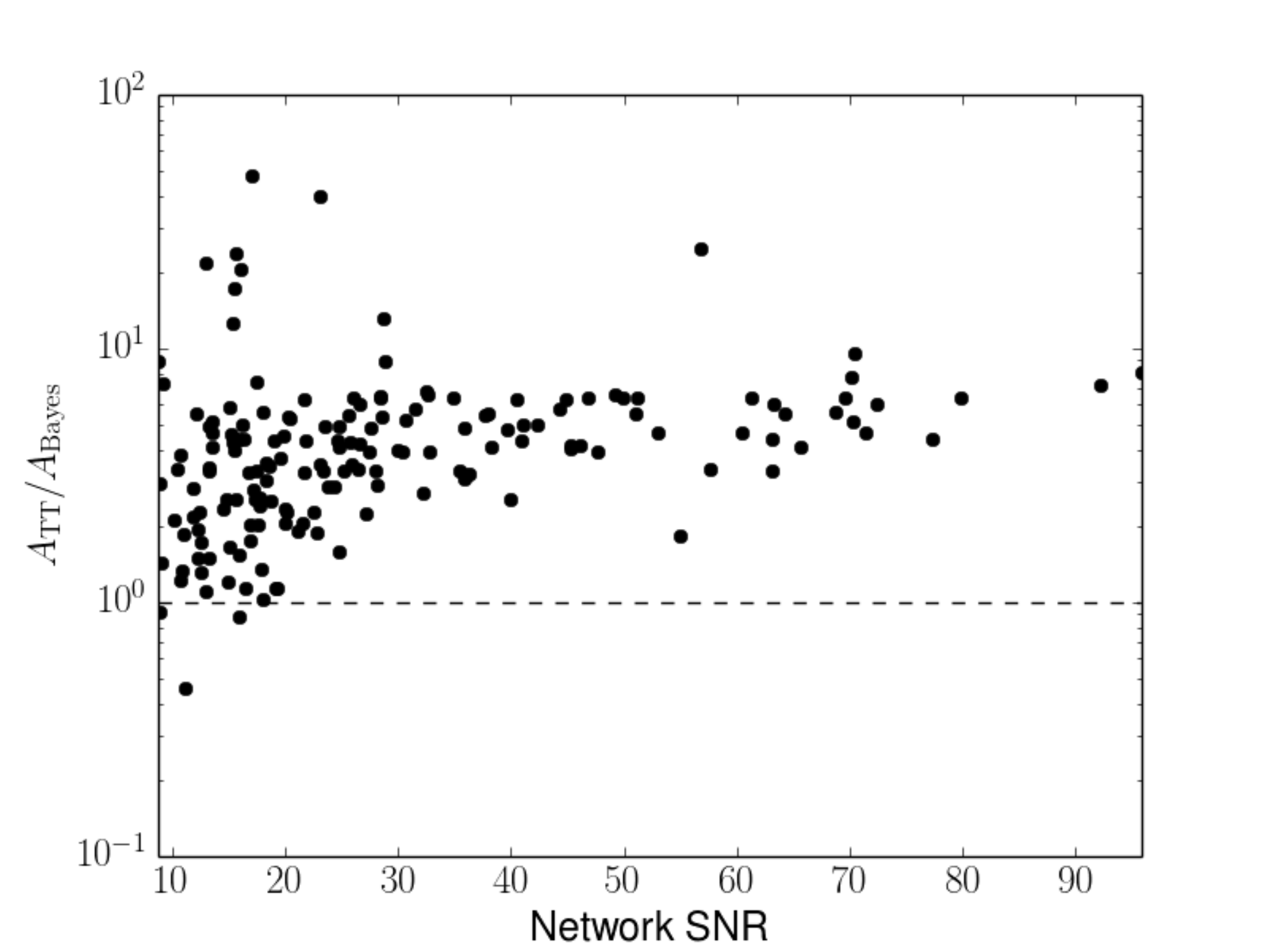}
	}%
	 \subfigure[$\,$9d Fisher information matrix and timing triangulation comparison ]{
		\label{fig:   fairhurst /fisher9}
		\includegraphics[width=0.49\textwidth]{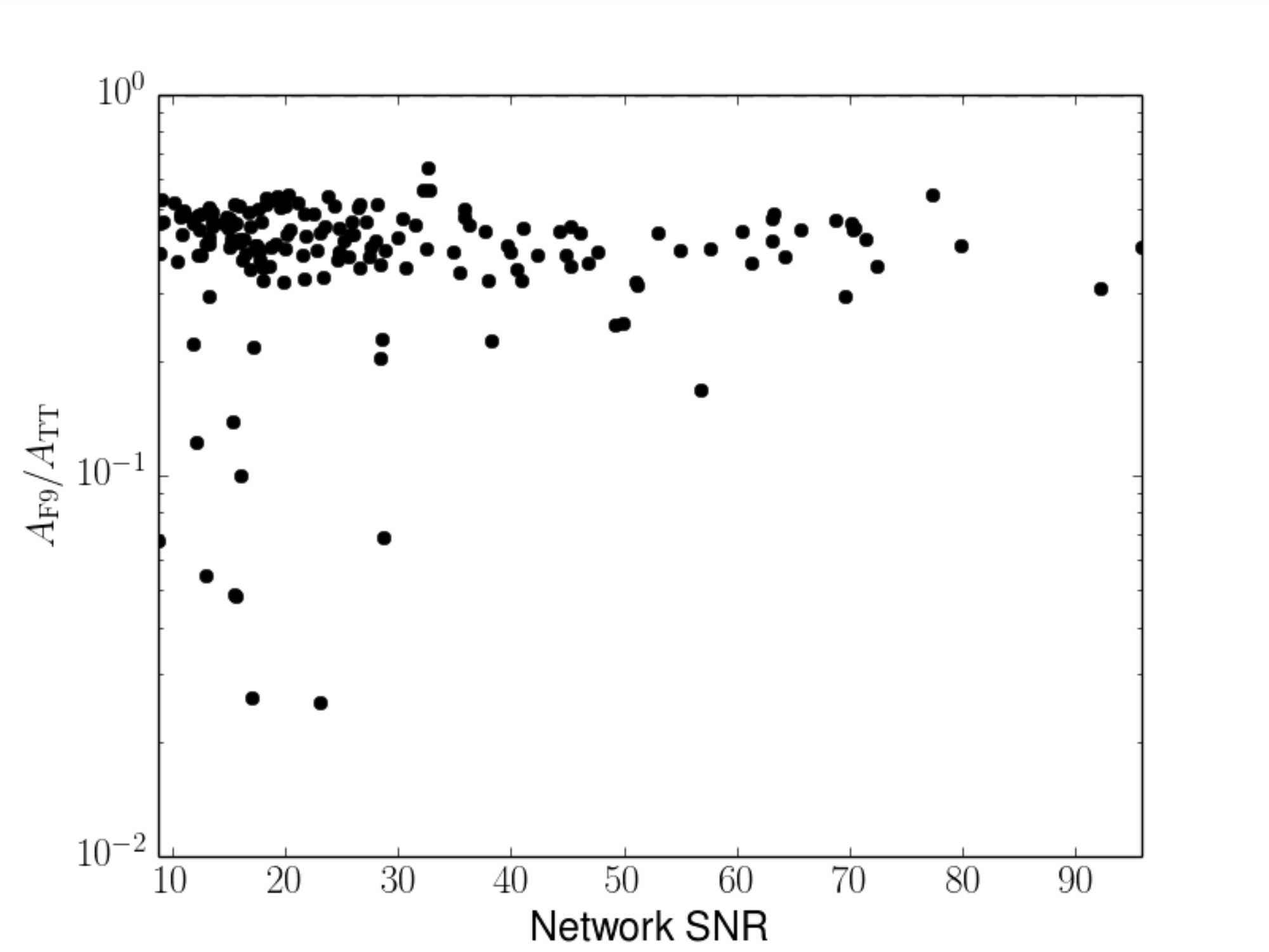}
	}\\
	\subfigure[$\,$9d Fisher information matrix and full coherent Bayesian comparison ]{
		\label{fig:   fisher 9d/mcmc}
		\includegraphics[width=0.49\textwidth]{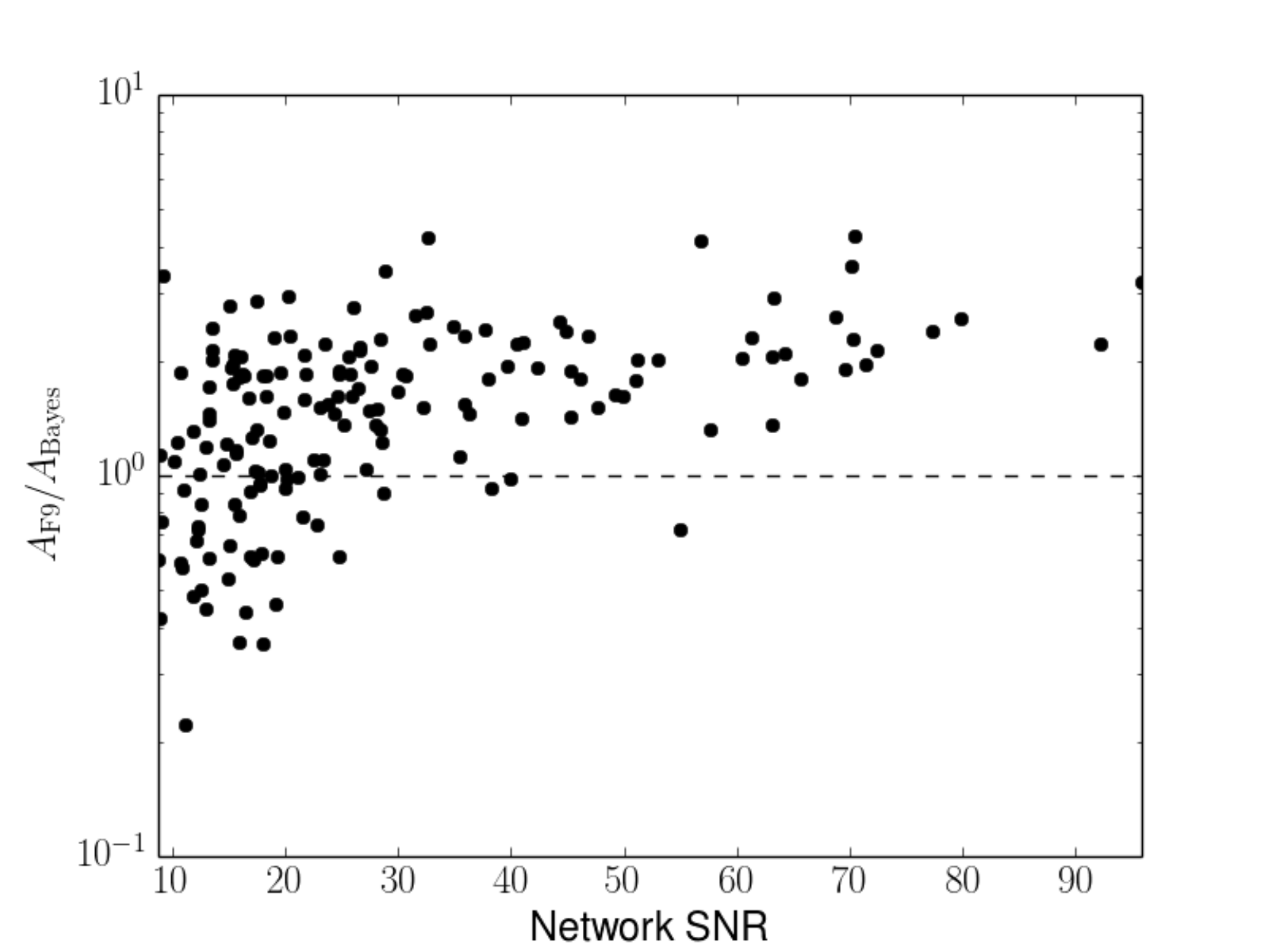}
	}%
	\subfigure[$\,$4d Fisher information matrix and full coherent Bayesian comparison ]{
		\label{fig:   fisher 4d/mcmc}
		\includegraphics[width=0.49\textwidth]{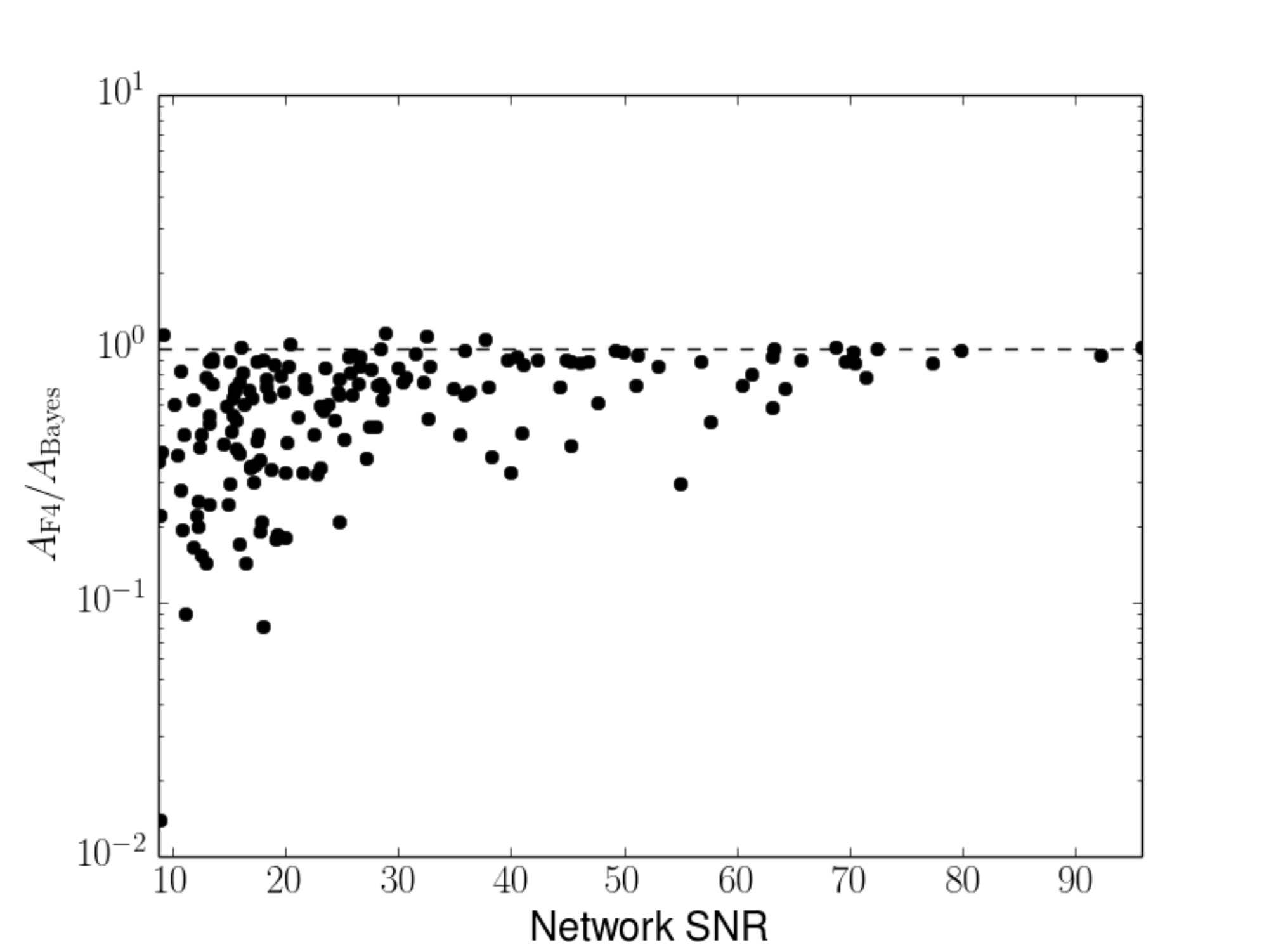}
	}%
	\caption{Comparisons of $50\%$ credible-interval sky areas found using the timing triangulation approximation $A_{\mathrm{TT}}$, the 4d Fisher information matrix approximation $A_{\mathrm{F4}}$, the 9d Fisher information matrix approximation $A_{\mathrm{F9}}$ and the coherent Bayesian method $A_{\mathrm{Bayes}}$, as a function of the network SNR of the signal.}
	\end{center}
\end{figure*}

The comparison of the timing triangulation (TT) approximate sky areas and those from the full Bayesian method provides an indication of the ability of the TT approximation to predict the sky localization uncertainty for GW transients with advanced networks.  It was expected that the extra information used in the full Bayesian coherent analysis should provide better localization and so smaller sky areas. However, prior to this study the size of this improvement was unknown. As seen in Table \ref{tab:summary}, the sky areas are on average significantly smaller when measured using the coherent Bayesian method  than when estimated with TT. When comparing  injections individually, the median of the ratio $A_{\mathrm{TT}} /A_{\mathrm{Bayes}}$ is  $4.1$, see Figure \ref{fig: fairhurst /mcmc}.  However, there is a large scatter in individual results.  The distribution of the ratio  $A_{\mathrm{TT}}/ A_{\mathrm{Bayes}}$ is strongly skewed, but the distribution of the logarithm of the ratio appears relatively Gaussian, so we use the standard deviation of the ratio as a proxy for the scatter.  The standard deviation of the natural logarithm of this ratio is $0.7$, meaning that individual sky localization areas typically differ by a factor of $\sim 2$.

\begin{figure*}[h]
	\begin{center}
        \subfigure[
        {$\,$Typical posterior density function}
        ]{
		\label{fig: inj126}
		\includegraphics[width=0.45\textwidth]{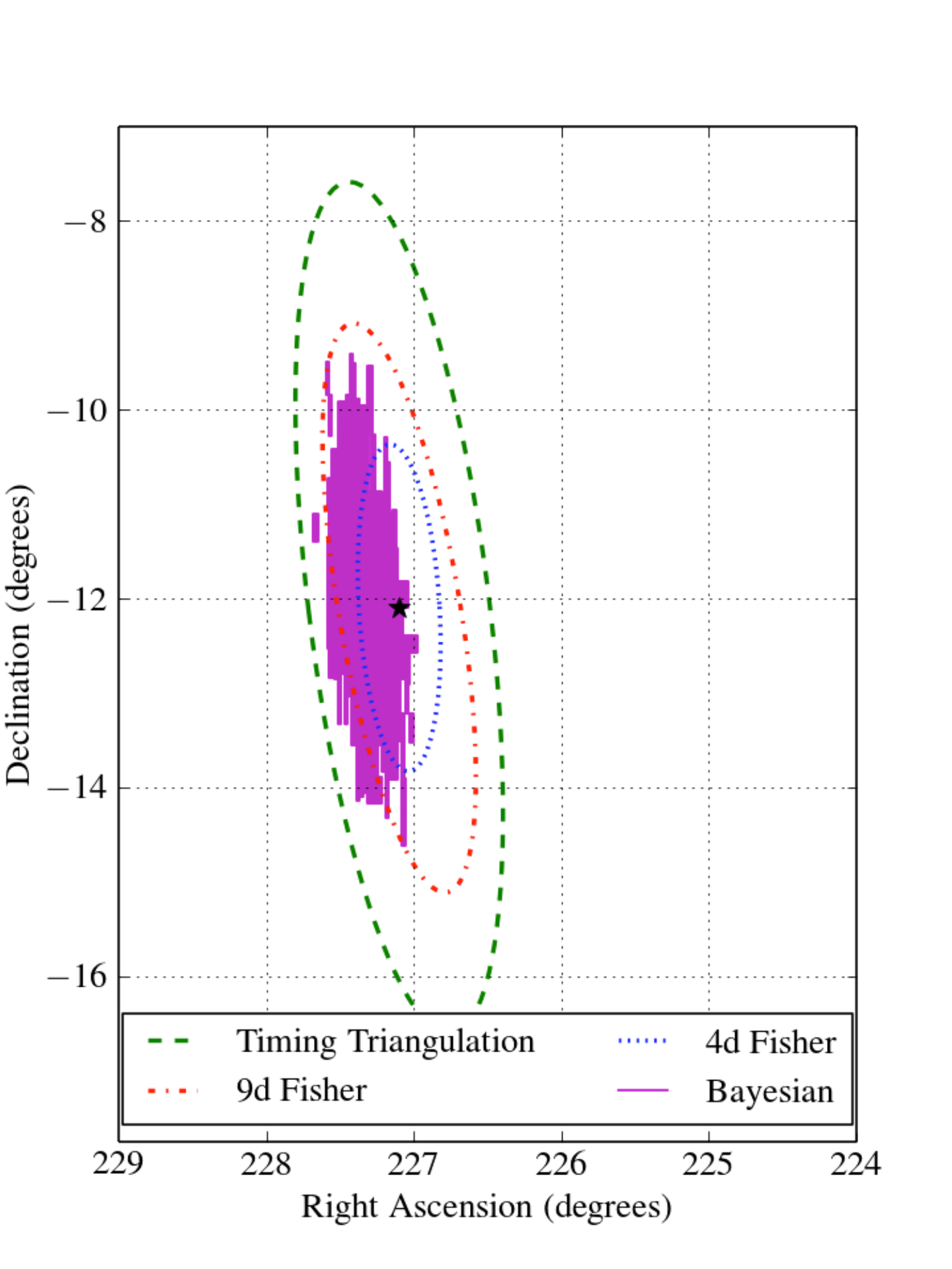}
		}
		\subfigure[
		{$\,$``Patchy'' posterior density function}
		]{
		\label{fig: inj190}
		\includegraphics[width=0.45\textwidth]{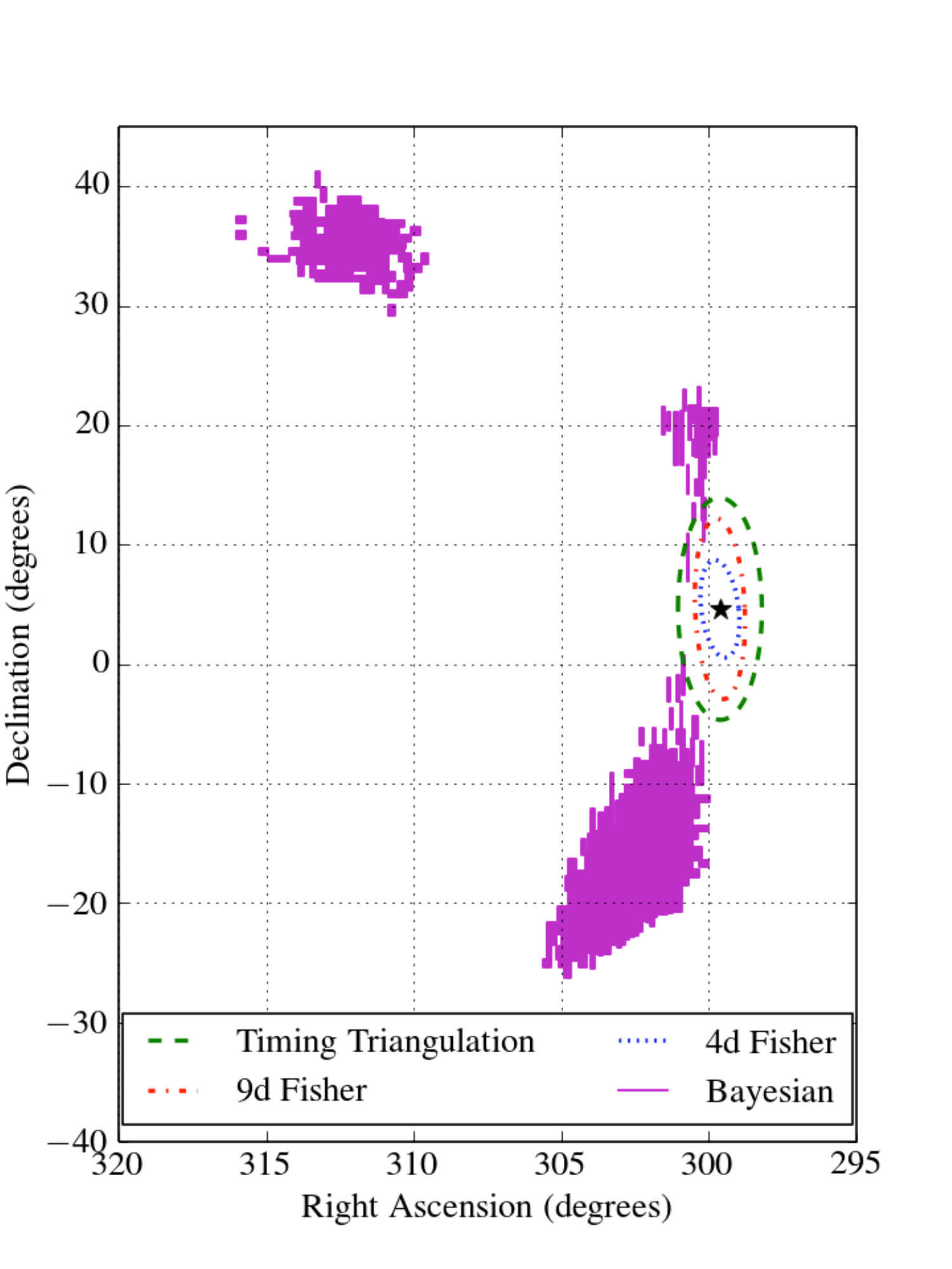}
		}
	\end{center}
	\caption{
	$50\%$ credible intervals for the a typical injection (panel (a), $A_\mathrm{TT}/ A_\mathrm{B} =4.01$, $A_\mathrm{F4}/A_\mathrm{} = 0.84$, $A_\mathrm{F9}/A_\mathrm{B} = 2.23$, $\rho_{\mathrm{Hanford}} = 7.48$, $\rho_{\mathrm{Livingston}} = 14.3$, $\rho_{\mathrm{Virgo}}= 17.1$ and  $\rho_{\mathrm{Network}}= 23.6$ ) and for a injection where the posterior density function found via the Bayesian analysis is patchy (panel (b), ${A_{\mathrm{TT}}}/ {A_{\mathrm{Bayes}}} = 0.38$, ${A_{\mathrm{F4}}}/ {A_{\mathrm{Bayes}}} = 0.09$, ${A_{\mathrm{F9}}}/ {A_{\mathrm{Bayes}}} = 0.22$, $\rho_{\mathrm{Hanford}} = 6.62$, $\rho_{\mathrm{Livingston}} = 4.54$, $\rho_{\mathrm{Virgo}}= 7.77$ and  $\rho_{\mathrm{Network}}= 11.2$). 
	} 
\end{figure*}

The Bayesian $50\%$ credible interval for a typical mock event is shown in Figure \ref{fig: inj126}, along with timing triangulation and FIM predictions for localization ellipses.  This event, at a network SNR of $23.6$, has $A_{\mathrm{TT}}/ A_{\mathrm{Bayes}} = 4.01$.  The Bayesian posteriors are slightly offset from the true value because of the impact of noise in the data, while the predicted TT and FIM uncertainties that do not consider the actual data realization are centered on the true value.  However, the qualitative shape of the Bayesian posterior here is ellipsoidal and the orientation and eccentricity of this ellipse match the analytical predictions, although the size of the ellipses vary because of the omission of key phase and amplitude information when constructing the TT uncertainty region.

Meanwhile, Figure \ref{fig: inj190} shows the Bayesian credible interval and the analytical localization predictions for the outlier seen in Figure \ref{fig:   fairhurst /mcmc} at a network SNR of $11.2$ and  $\frac{A_{\mathrm{TT}}} {A_{\mathrm{Bayes}}} =0.38$.   In this case, the partial degeneracy between multiple sky locations cannot be broken, and the Bayesian $50\%$ credible interval is composed of multiple patches in the sky. The TT and FIM technique, which assume that such degeneracies can always be broken using amplitude information, do not account for this degeneracy and significantly under-predict the sky localization uncertainty.

\begin{figure*}[h]
	\begin{center}
	\subfigure[$\, 30\%$ Credible Interval ]{
		\label{fig: gaussian 30}
		\includegraphics[width=0.49\textwidth]{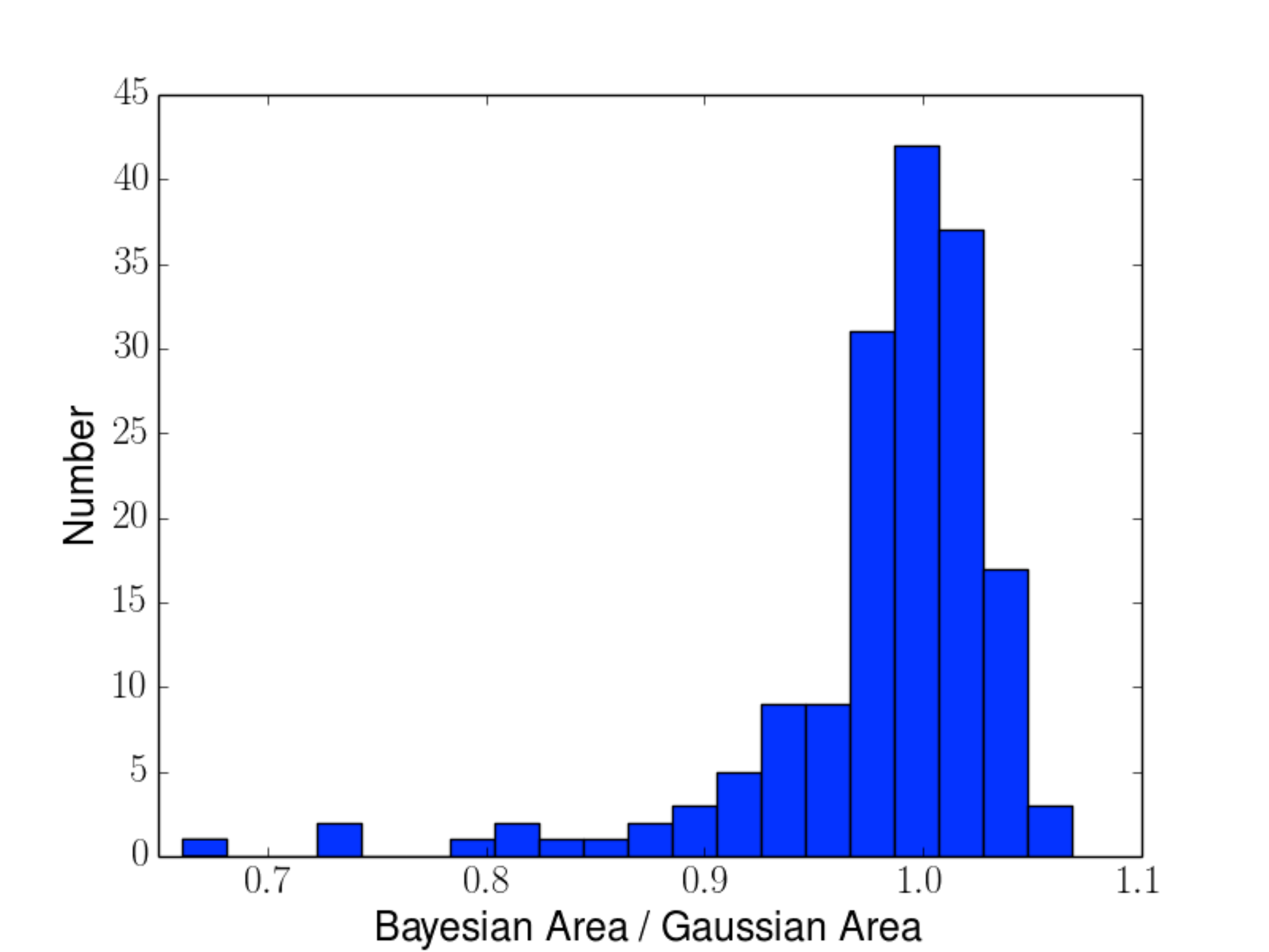}
		}%
	\subfigure[$\, 70\%$ Credible Interval]{
		\label{fig: gaussian 70}
		\includegraphics[width=0.49\textwidth]{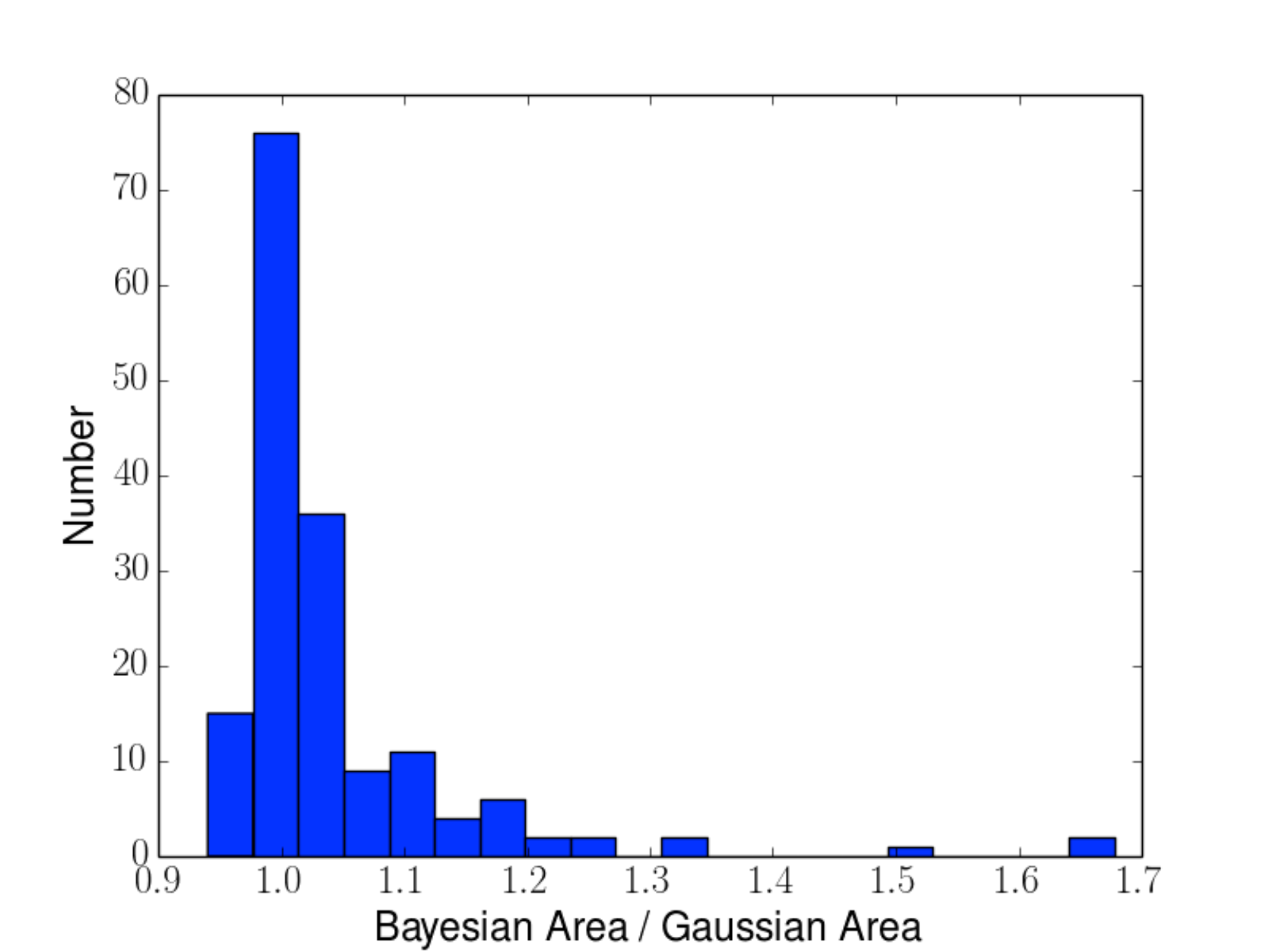}
		}%
	\end{center}
	\caption{Histograms of Bayesian sky areas at the $30\%$ and $70\%$ credible levels to the area of a Gaussian posterior at the same credible level, scaled to match the Bayesian posterior at the $50\%$ credible level.}
	\label{fig:subfigures gauss }
\end{figure*}

As the TT and FIM approximations assume a Gaussian posterior distribution, any deviations from this, as seen in an extreme case in Figure \ref{fig: inj190}, will affect the validity of these approximations. In order to investigate the Gaussianity of the Bayesian posteriors we compare the two-dimensional $30\%$ credible intervals and $70\%$ credible intervals to those expected for a Gaussian posterior distribution scaled to match the Bayesian posterior at the $50\%$ CL, Figure \ref{fig:subfigures gauss }. The tails extending to lower ratios of Bayesian to Gaussian ratios at the $30\%$ credible intervals and to higher ratios at the $70\%$ credible intervals CI indicate that some of the Bayesian posteriors are not Gaussian, but have greater peakedness with long tails.

\subsection{Fisher Matrix vs Coherent Bayesian}

The fractional difference in the sky location angles, right ascension and declination, between FIM and Bayesian for this set of simulated sources have been reported in \cite{rodriguez13}, where it was shown that the FIM tends to overestimate the error in individual sky angles at low SNRs and underestimate it at high SNRs. In Figure \ref{fig:   fairhurst /fisher9} we show the ratio of sky areas for the FIM and TT approximations, while Figure \ref{fig:   fisher 9d/mcmc} shows the ratio of FIM sky areas to the coherent Bayesian $50\%$ credible intervals.  With the cuts used in this study (see Section \ref{sec:det_inj}), the median of ratios $A_{F9}/A_{Bayes}$ is $1.6$ with standard deviation in the log of the ratio $0.6$, while the median of the ratio $A_{TT}/A_{F9}$ is  $2.4$  and standard deviation in the log of the ratio of $0.5$. The inclusion of the full 9-dimensional parameter space in the FIM results provides a closer estimation of the full coherent analysis than the areas estimated via timing triangulation alone, but still overestimates the sky areas.  While the FIM does include all the signal parameters, it is still limited to considering the leading order, quadratic terms to determine localization, and it makes no use of physically motivated priors on the signal parameters.  Both of these are included in the Bayesian approach and are likely to explain the difference in results.

The restricted 4-dimensional FIM analysis over-constrains the sky areas by artificially assuming perfect knowledge of the remaining 5 parameters.  The ratio of 4-dimensional FIM $50\%$ confidence-interval sky areas to corresponding areas from the coherent Bayesian analysis $A_{F4}/A_{\mathrm{Bayes}}$, shown in Figure \ref{fig:   fisher 4d/mcmc},  have a median $0.68$ with standard deviation in the log of the ratio $0.6$.  It is not surprising that the 4-dimensional FIM gives the smallest localization regions, as it is imposing physically unrealistic restrictions on the parameters.

\subsection{Timing Triangulation with Phase Correction \label{TTphase}}

We have seen that signal parameters not considered in basic timing triangulation still affect the sky localization, yielding an area that is a factor of $2.7$ larger than predicted by the 9-dimensional FIM. Ignoring these parameters, particularly the phase consistency between detectors, leads to pessimistic predictions of the ability of a detector network to localize GW  sources. We can attempt to correct the predicted sky localization based on timing triangulation alone \cite{fairhurst2009,fairhurst2011} to account for phase information.   

To understand the impact of this, we can consider a simplified case where we ignore the polarization of the waveform and keep the relative phase of the signal between detectors fixed.  Generically, this is overly restrictive, although it would be appropriate for a face-on, circularly polarized binary.  

Consider a waveform of the form $\tilde{h}(f) = A(f) \exp\left[{i \phi_0 + i 2\pi f t_0 + i \Psi(f)}\right]$ where $\phi_0$ and $t_0$ are the phase and time at a particular frequency (e.g., at the fiducial moment of coalescence) and $A(f)$ and $\Psi(f)$ are the amplitude and phase, respectively, which depend on the other seven parameters of the signal. For a signal of this form, it is possible to measure both $\phi_0$ and $t_0$ from each detector's data.  The timing accuracy in each detector $\sigma_t$, considered independently, can be obtained by applying a 2-dimensional FIM calculation in $\{\phi_0,t_0\}$ to each detector's data set. This FIM is:
\begin{eqnarray}
\Gamma_{tt} &=& 4\pi^2 \rho^2 \overline{f^2}\, ,\nonumber\\
\Gamma_{t\phi} &=& 2 \pi \rho^2 \bar{f}\, ,\\ 
\Gamma_{\phi\phi} &=& \rho^2\, , \nonumber
\end{eqnarray}
where $\bar{f}$ and $\overline{f^2}$ are defined in Eq.~(\ref{eqn: fbar}).
In the high SNR limit, the covariance matrix can be approximated by the inverse of the FIM, Eq.~(\ref{eqn:covar}).  The timing uncertainty in an individual detector is given by $\sigma_t^2 =\left( \Gamma^{-1}\right)_{tt} = \left[4 \pi^{2}\rho^{2} (\overline{f^2} - \bar{f}^2)\right]^{-1}$, in agreement with the expression for $\sigma_t$ given in Eq.~(\ref{eqn: sigma t}).

This calculation, a variant of which was used for the original TT prediction \cite{fairhurst2009}, ignores the requirement of phase consistency between detectors.  
Although we donÕt know the actual value of $\phi_0$ in the detectors, its value must be the same for all detectors regardless of detector location. 
Fixing the phase, rather than marginalizing over it, corresponds to using a 1-dimensional FIM $\Gamma_{tt}$ in place of the two-dimensional matrix considered above.  Thus, with the phase consistency requirement, the timing uncertainty in each detector is just the inverse of the $tt$ component of the 2-dimensional FIM, $\sigma_{\textrm t,\,\mathrm{new}}^2 = \left(\Gamma_{tt}\right)^{-1}$.   We can use this timing uncertainty in place of $\sigma_t$ in Eq.~(\ref{eq sky loc matrix}) to compute a prediction for the sky localization uncertainty with the phase consistency requirement.

\begin{figure}[h]
	\includegraphics[width=0.49\textwidth]{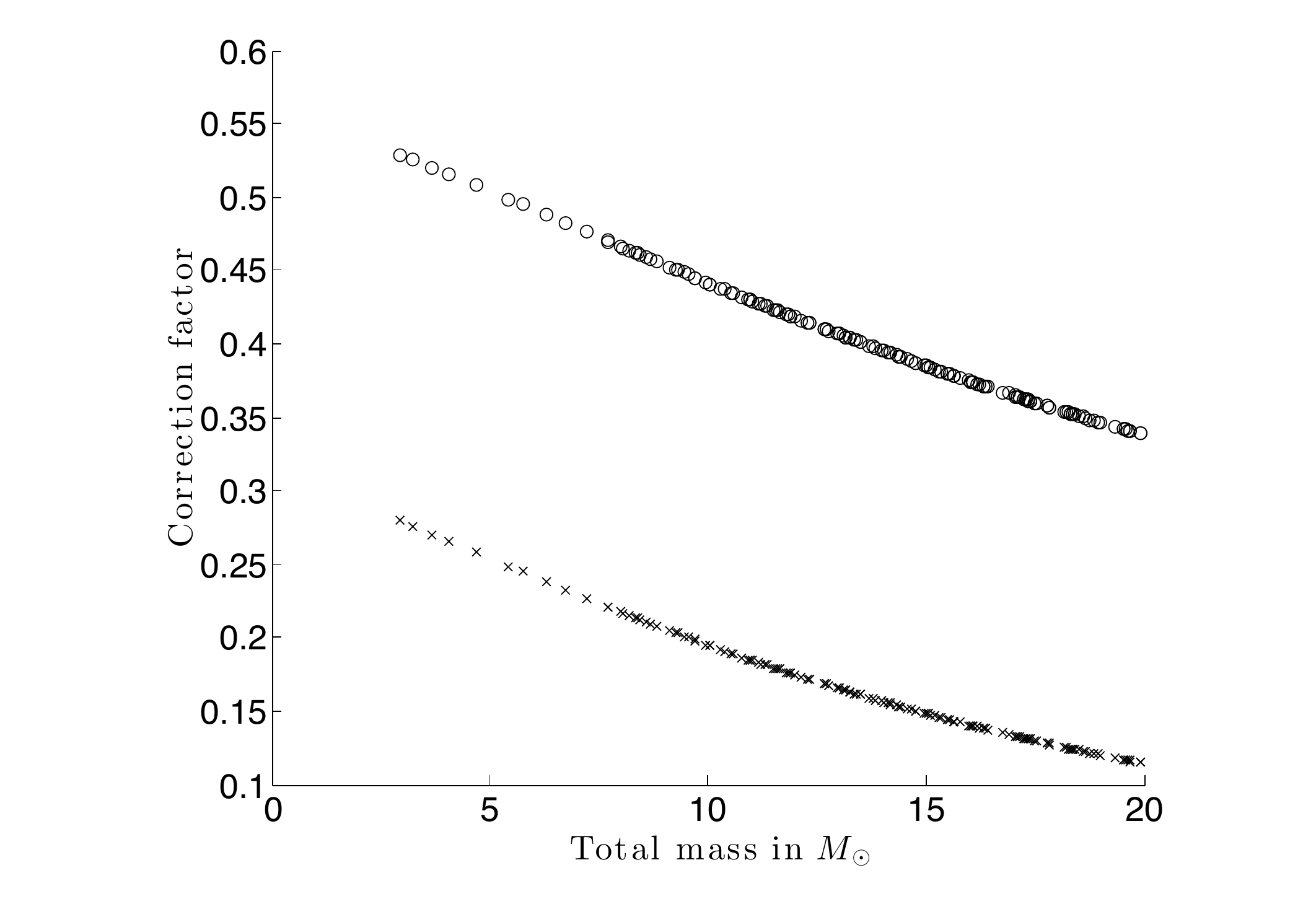}
        \caption{Fractional reduction in timing triangulation prediction for the sky localization area after incorporating phase consistency versus using only timing information;  crosses correspond to the single-free-phase approximation described here, while circles correspond to the two-free-phase approximation from \cite{fairhurst2013}. }
        	\label{fig:correction}
\end{figure}

If there is a correlation between $\phi_0$ and $t_0$, the uncertainty ellipse in time--phase space will be inclined.  In this case, the ellipse's projection onto the time axis, measured by $\sigma_t$, which was the measure of localization used in the original TT approach, can be much wider than the width of the ellipse at a fixed value of phase.   Thus, the largest improvements from including phase constraints in timing triangulation arise when there are large correlations between time and phase measurements.  For the simple case of detectors with equal noise power spectral densities, such as those considered in this study, sky localization area uncertainties will decrease by a factor of  
\begin{equation}
 \frac{\sigma_{t,\,\textrm{new}}^2}{\sigma_t^2} = \frac{\overline{f^2}-\bar{f}^2}{\overline{f^2}}\, .
\end{equation}
The precise value added depends on both the detector bandwidth and the system masses, with larger contributions for higher masses because of greater correlation (see Figure \ref{fig:correction}).  
For instance, for a signal from a typical binary neutron star system observed by the network considered here, we expect approximately a factor of $3$ improvement in sky localization by including phasing information. 
However, for the most massive systems in the set of injections considered above, the improvement could be as large as a factor of $8$.  

Polarization is treated more carefully in \cite{fairhurst2013}.  The additional freedom contained in the choice of polarization lowers the phase consistency improvement to the sky localization area to only a square root of the value predicted above, i.e.,
\begin{equation}
  \frac{\sigma_{t,\, \textrm{new}}^2}{\sigma_t^2} = \sqrt {\frac{\overline{f^2}-\bar{f}^2}{\overline{f^2}}}\, .
\end{equation}
Heuristically, this can be understood by noting that the GW has two free phases, while TT (with a three-site network) assumes three and the derivation above restricted to a single free phase.  Thus, we should expect the actual result to be midway between TT and fixed phase approximations.  Folding in phasing information in this way, the improvement in predicted sky localization relative to TT alone is a factor of $1.8$ for a signal from a typical binary neutron star system, and as much as a factor of three for the signal from the most massive system in the set of injections considered above.
Note, however, that the effect will be less significant in advanced detectors as their sensitive band starts at a lower frequency.

%
%

\begin{table}[h]
	\begin{center}
	\begin{tabular}{ccc}
		\hline
		\hline
		Method & 	Median Area  & Median Ratio  \\
		&  (square degrees) & with $A_{\mathrm{Bayes}}$ \\
		\hline
		Bayesian  &  2.9 &  - \\
		standard TT & 10.6 & 4.1 \\
		TT + two phases from \cite{fairhurst2013} & 4.4 & 1.6\\
		TT + one free phase  & 1.7 & 0.6 \\
 		\hline
		\hline
	\end{tabular}
	\end{center}
	\caption{The median areas of $50\%$ Bayesian credible intervals for coherent Bayesian method, timing triangulation and time-phase localization. Median ratios of the areas relative to the coherent Bayesian method are also given.}
	\label{tab:Median new TT}
\end{table}

We see in Table \ref{tab:Median new TT} that the inclusion of phase information improves the TT predictions and brings them closer in-line with the observed sky localization ability of coherent Bayesian analysis.  In the case where we ignore the freedom of choosing the polarization angle, we effectively over-constrained the system with the phase consistency requirement, which leads to predicted credible intervals that are smaller than what can be achieved even with coherent Bayesian analysis.  In fact, the median $50\%$ credible-interval area for the TT prediction with this phase constraint, $1.7$ square degrees for our injection set  (see Table \ref{tab:Median new TT}), matches with the 4-dimensional FIM median interval of $1.6$ square degrees (see Table\ref{tab:summary}).  This is to be expected as both methods effectively consider three free angles: two sky angles and phase.

Predictions from the TT calculation with two free phases are similar to the full 9-dimensional FIM, and, indeed, the median sky localization areas for the two methods are $4.4$ and $3.9$ square degrees, respectively.  This indicates that by requiring a coherent signal across the detectors, we have incorporated the most significant additional parameter after arrival time.  However, the median area predicted with this method is more than a factor of two larger than the median Bayesian area, which indicates that additional inputs can further improve sky localization ability.

This importance of coherence, and particularly phase consistency, is relevant for the development of actual (rather than merely predictive) rapid methods to localize sources based on TT, e.g., \cite{abadie12}.  We expect these rapid localization methods to benefit from including phase consistency information along with timing triangulation.  As discussed above, the largest benefits accrue for more massive systems.  

Although we see that the inclusion of phase information brings average predictions of timing-triangulation based estimates closer to the full coherent Bayesian localization, there is very significant scatter in results on individual mock data sets when compared on an injection-by-injection basis.  The standard deviation of the natural logarithm of the ratio of the $50\%$ MCMC credible intervals to $50\%$ TT predicted areas is $\sim 0.7$ both without phase information and for both correction schemes.   

\section{Discussion \label{sec:discussion}}

We have compared the predicted accuracy of source localization with various techniques with the performance obtained by a fully coherent Bayesian parameter estimation method.  In general, the full exploration of the parameter space yields the most accurate results, and also the smallest confidence regions.  For the systems considered here, the median areas are a factor of four smaller than those obtained with the timing triangulation approximation.  We have seen that requiring a consistent phase between the sites can significantly improve the localization of sources.  The agreement between this result and the 9-dimensional FIM result suggests that this is the most significant additional parameter to include.  For the set of simulations considered here, incorporating this correlation reduces the median area by a factor of $2.5$.  However, for low-mass signals the correction will be on the order of 50\%; the inclusion of merger and ringdown waveform phases will similarly reduce the impact of phase consistency for stellar-mass black hole binaries.

The best of the approximation methods still differ, on average, by at least 50\% from the full results and, more significantly, there is a large variation in the results on an event-by-event basis.  The difference between the full analysis results and the best approximations indicate that there are still significant factors which are being overlooked in the approximation methods.  These include: correlations with other parameters such as the binary's component masses, priors on signal parameters, particularly distance and inclination, and the specific noise realization in the data set being analyzed. Furthermore, all of the approximation techniques consider only the leading-order effects.  At low SNR, expected for the first gravitational-wave detections, these approximations are not sufficient and a full examination of the parameter space is required to accurately extract the signal parameters.  While the approximation methods can be used to give a sense of the localization capabilities of various detector networks, the fully coherent Bayesian analysis gives the most accurate results both on a population of sources and, most significantly, on individual events.  It also provides a critical cross-check on the self-consistency of purely predictive methods techniques for parameter estimation.

In this study, we only analyzed the case of non-spinning compact objects.  The results should extend straightforwardly to systems with aligned spins.  However, for binaries in which at least one of the components is a black hole, and therefore could carry significant spin, precession of the orbital plane through spin-orbit and spin-spin coupling carries information about the system location and orientation parameters and therefore affects our ability to recover the sky location of the source. It is likely that these effects will produce a systematic offset in a parameter estimation approach based on intermediate data products from single-detector analyses.  A fully coherent Bayesian parameter recovery can straightforwardly incorporate such effects. It would be interesting to investigate the effect of precession on source localization, but we leave that for the future.

\section*{Acknowledgements}
We would like to acknowledge the useful discussions with our colleges from the LIGO -- Virgo Collaboration, especially the helpful comments from Leo Singer. This work has been supported in part by the UK Science and Technology Facilities Council.  SF would like to thank the Royal Society for support.

\bibliography{SkyLocPaperDraft}

\begin{thebibliography}{30}
\expandafter\ifx\csname natexlab\endcsname\relax\def\natexlab#1{#1}\fi
\expandafter\ifx\csname bibnamefont\endcsname\relax
  \def\bibnamefont#1{#1}\fi
\expandafter\ifx\csname bibfnamefont\endcsname\relax
  \def\bibfnamefont#1{#1}\fi
\expandafter\ifx\csname citenamefont\endcsname\relax
  \def\citenamefont#1{#1}\fi
\expandafter\ifx\csname url\endcsname\relax
  \def\url#1{\texttt{#1}}\fi
\expandafter\ifx\csname urlprefix\endcsname\relax\def\urlprefix{URL }\fi
\providecommand{\bibinfo}[2]{#2}
\providecommand{\eprint}[2][]{\url{#2}}

\bibitem[{\citenamefont{{Harry} and {the LIGO Scientific
  Collaboration}}(2010)}]{AdvLIGO}
\bibinfo{author}{\bibfnamefont{G.~M.} \bibnamefont{{Harry}}} \bibnamefont{and}
  \bibinfo{author}{\bibnamefont{{the LIGO Scientific Collaboration}}},
  \bibinfo{journal}{Class. Quantum Gravity} \textbf{\bibinfo{volume}{27}},
  \bibinfo{pages}{084006} (\bibinfo{year}{2010}), \eprint{arXiv:1103.2728}.

\bibitem[{\citenamefont{{Virgo Collaboration}}(2009)}]{AdvVirgo}
\bibinfo{author}{\bibnamefont{{Virgo Collaboration}}}, \bibinfo{type}{Virgo
  Technical Report} \bibinfo{number}{VIR-0027A-09} (\bibinfo{year}{2009}),
  \bibinfo{note}{https://tds.ego-gw.it/itf/tds/file.php?callFile=VIR-0027A-09.pdf}.

\bibitem[{\citenamefont{{ {Abadie}, J. et al. ({the LIGO Scientific
  Collaboration} and {the Virgo Collaboration})}}(2010)}]{LVC2010}
\bibinfo{author}{\bibnamefont{{ {Abadie}, J. et al. ({the LIGO Scientific
  Collaboration} and {the Virgo Collaboration})}}}, \bibinfo{journal}{Class.
  Quantum Gravity} \textbf{\bibinfo{volume}{27}}, \bibinfo{pages}{173001}
  (\bibinfo{year}{2010}), \eprint{arXiv:1003.2480}.

\bibitem[{\citenamefont{{Kalogera} et~al.}(2007)\citenamefont{{Kalogera},
  {Belczynski}, {Kim}, {O'Shaughnessy}, and {Willems}}}]{Kalogera:2007}
\bibinfo{author}{\bibfnamefont{V.}~\bibnamefont{{Kalogera}}},
  \bibinfo{author}{\bibfnamefont{K.}~\bibnamefont{{Belczynski}}},
  \bibinfo{author}{\bibfnamefont{C.}~\bibnamefont{{Kim}}},
  \bibinfo{author}{\bibfnamefont{R.}~\bibnamefont{{O'Shaughnessy}}},
  \bibnamefont{and}
  \bibinfo{author}{\bibfnamefont{B.}~\bibnamefont{{Willems}}},
  \bibinfo{journal}{Phys. Rep.} \textbf{\bibinfo{volume}{442}},
  \bibinfo{pages}{75} (\bibinfo{year}{2007}), \eprint{arXiv:astro-ph/0612144}.

\bibitem[{\citenamefont{{Mandel} and
  {O'Shaughnessy}}(2010)}]{MandelOShaughnessy:2010}
\bibinfo{author}{\bibfnamefont{I.}~\bibnamefont{{Mandel}}} \bibnamefont{and}
  \bibinfo{author}{\bibfnamefont{R.}~\bibnamefont{{O'Shaughnessy}}},
  \bibinfo{journal}{Class. Quantum Gravity} \textbf{\bibinfo{volume}{27}},
  \bibinfo{pages}{114007} (\bibinfo{year}{2010}), \eprint{arXiv:0912.1074}.

\bibitem[{\citenamefont{{Li} et~al.}(2012)\citenamefont{{Li}, {Del Pozzo},
  {Vitale}, {Van Den Broeck}, {Agathos}, {Veitch}, {Grover}, {Sidery},
  {Sturani}, and {Vecchio}}}]{Li:2012}
\bibinfo{author}{\bibfnamefont{T.~G.~F.} \bibnamefont{{Li}}},
  \bibinfo{author}{\bibfnamefont{W.}~\bibnamefont{{Del Pozzo}}},
  \bibinfo{author}{\bibfnamefont{S.}~\bibnamefont{{Vitale}}},
  \bibinfo{author}{\bibfnamefont{C.}~\bibnamefont{{Van Den Broeck}}},
  \bibinfo{author}{\bibfnamefont{M.}~\bibnamefont{{Agathos}}},
  \bibinfo{author}{\bibfnamefont{J.}~\bibnamefont{{Veitch}}},
  \bibinfo{author}{\bibfnamefont{K.}~\bibnamefont{{Grover}}},
  \bibinfo{author}{\bibfnamefont{T.}~\bibnamefont{{Sidery}}},
  \bibinfo{author}{\bibfnamefont{R.}~\bibnamefont{{Sturani}}},
  \bibnamefont{and}
  \bibinfo{author}{\bibfnamefont{A.}~\bibnamefont{{Vecchio}}},
  \bibinfo{journal}{Phys. Rev. D} \textbf{\bibinfo{volume}{85}},
  \bibinfo{pages}{082003} (\bibinfo{year}{2012}), \eprint{arXiv:1110.0530}.

\bibitem[{\citenamefont{{Metzger} and {Berger}}(2012)}]{metzger2012}
\bibinfo{author}{\bibfnamefont{B.~D.} \bibnamefont{{Metzger}}}
  \bibnamefont{and} \bibinfo{author}{\bibfnamefont{E.}~\bibnamefont{{Berger}}},
  \bibinfo{journal}{The Astrophysical Journal} \textbf{\bibinfo{volume}{746}},
  \bibinfo{pages}{48} (\bibinfo{year}{2012}), \eprint{arXiv:1108.6056}.

\bibitem[{\citenamefont{{Abadie}}(2012)}]{abadie12}
\bibinfo{author}{\bibfnamefont{J.~e.~a.} \bibnamefont{{Abadie}}},
  \bibinfo{journal}{Astron. Astrophys.} \textbf{\bibinfo{volume}{541}},
  \bibinfo{pages}{A155} (\bibinfo{year}{2012}), \eprint{arXiv:1112.6005}.

\bibitem[{\citenamefont{{{Aasi}, J. et al. (LIGO-Virgo ScientiÞc
  Collaboration)}}(2013)}]{LVC2013a}
\bibinfo{author}{\bibnamefont{{{Aasi}, J. et al. (LIGO-Virgo ScientiÞc
  Collaboration)}}}, \bibinfo{journal}{Phys. Rev. D}
  \textbf{\bibinfo{volume}{88}}, \bibinfo{pages}{062001}
  (\bibinfo{year}{2013}), \eprint{arXiv:1304.1775}.

\bibitem[{\citenamefont{{van der Sluys} et~al.}(2008)\citenamefont{{van der
  Sluys}, {Raymond}, {Mandel}, {R{\"o}ver}, {Christensen}, {Kalogera}, {Meyer},
  and {Vecchio}}}]{vanderSluys2008}
\bibinfo{author}{\bibfnamefont{M.}~\bibnamefont{{van der Sluys}}},
  \bibinfo{author}{\bibfnamefont{V.}~\bibnamefont{{Raymond}}},
  \bibinfo{author}{\bibfnamefont{I.}~\bibnamefont{{Mandel}}},
  \bibinfo{author}{\bibfnamefont{C.}~\bibnamefont{{R{\"o}ver}}},
  \bibinfo{author}{\bibfnamefont{N.}~\bibnamefont{{Christensen}}},
  \bibinfo{author}{\bibfnamefont{V.}~\bibnamefont{{Kalogera}}},
  \bibinfo{author}{\bibfnamefont{R.}~\bibnamefont{{Meyer}}}, \bibnamefont{and}
  \bibinfo{author}{\bibfnamefont{A.}~\bibnamefont{{Vecchio}}},
  \bibinfo{journal}{Class. Quantum Gravity} \textbf{\bibinfo{volume}{25}},
  \bibinfo{pages}{184011} (\bibinfo{year}{2008}), \eprint{arXiv:0805.1689}.

\bibitem[{\citenamefont{Veitch and Vecchio}(2010)}]{veitch10}
\bibinfo{author}{\bibfnamefont{J.}~\bibnamefont{Veitch}} \bibnamefont{and}
  \bibinfo{author}{\bibfnamefont{A.}~\bibnamefont{Vecchio}},
  \bibinfo{journal}{Phys. Rev. D} \textbf{\bibinfo{volume}{81}},
  \bibinfo{pages}{062003} (\bibinfo{year}{2010}), \eprint{arXiv:0911.3820}.

\bibitem[{\citenamefont{{LSC Algorithm Library software}}()}]{LAL}
\bibinfo{author}{\bibnamefont{{LSC Algorithm Library software}}},
  \emph{\bibinfo{title}{{\tt http://www.lsc-group.phys.uwm.edu/lal}}}.

\bibitem[{\citenamefont{Veitch et~al.}(2012)\citenamefont{Veitch, Mandel,
  Aylott, Farr, Raymond, Rodriguez, van~der Sluys, Kalogera, and
  Vecchio}}]{veitch2012}
\bibinfo{author}{\bibfnamefont{J.}~\bibnamefont{Veitch}},
  \bibinfo{author}{\bibfnamefont{I.}~\bibnamefont{Mandel}},
  \bibinfo{author}{\bibfnamefont{B.}~\bibnamefont{Aylott}},
  \bibinfo{author}{\bibfnamefont{B.}~\bibnamefont{Farr}},
  \bibinfo{author}{\bibfnamefont{V.}~\bibnamefont{Raymond}},
  \bibinfo{author}{\bibfnamefont{C.}~\bibnamefont{Rodriguez}},
  \bibinfo{author}{\bibfnamefont{M.}~\bibnamefont{van~der Sluys}},
  \bibinfo{author}{\bibfnamefont{V.}~\bibnamefont{Kalogera}}, \bibnamefont{and}
  \bibinfo{author}{\bibfnamefont{A.}~\bibnamefont{Vecchio}},
  \bibinfo{journal}{Phys. Rev. D} \textbf{\bibinfo{volume}{85}},
  \bibinfo{pages}{104045} (\bibinfo{year}{2012}), \eprint{arXiv:1201.1195}.

\bibitem[{\citenamefont{{Nissanke} et~al.}(2013)\citenamefont{{Nissanke},
  {Kasliwal}, and {Georgieva}}}]{Nissanke:2013}
\bibinfo{author}{\bibfnamefont{S.}~\bibnamefont{{Nissanke}}},
  \bibinfo{author}{\bibfnamefont{M.}~\bibnamefont{{Kasliwal}}},
  \bibnamefont{and}
  \bibinfo{author}{\bibfnamefont{A.}~\bibnamefont{{Georgieva}}},
  \bibinfo{journal}{Astrophys. J.} \textbf{\bibinfo{volume}{767}},
  \bibinfo{eid}{124} (\bibinfo{year}{2013}), \eprint{arXiv:1210.6362}.

\bibitem[{\citenamefont{{Kasliwal} and {Nissanke}}(2013)}]{kasliwal13}
\bibinfo{author}{\bibfnamefont{M.~M.} \bibnamefont{{Kasliwal}}}
  \bibnamefont{and}
  \bibinfo{author}{\bibfnamefont{S.}~\bibnamefont{{Nissanke}}},
  \bibinfo{journal}{ArXiv e-prints}  (\bibinfo{year}{2013}),
  \eprint{arXiv:1309.1554}.

\bibitem[{\citenamefont{Cutler and Flanagan}(1994)}]{cutler1994}
\bibinfo{author}{\bibfnamefont{C.}~\bibnamefont{Cutler}} \bibnamefont{and}
  \bibinfo{author}{\bibfnamefont{E.~E.} \bibnamefont{Flanagan}},
  \bibinfo{journal}{Phys. Rev. D} \textbf{\bibinfo{volume}{49}},
  \bibinfo{pages}{2658} (\bibinfo{year}{1994}), \eprint{arXiv:gr-qc/9402014}.

\bibitem[{\citenamefont{{Vallisneri}}(2008)}]{Vallisneri:2008}
\bibinfo{author}{\bibfnamefont{M.}~\bibnamefont{{Vallisneri}}},
  \bibinfo{journal}{Phys. Rev. D} \textbf{\bibinfo{volume}{77}},
  \bibinfo{pages}{042001} (\bibinfo{year}{2008}), \eprint{arXiv:gr-qc/0703086}.

\bibitem[{\citenamefont{{Rodriguez}
  et~al.}(2013{\natexlab{a}})\citenamefont{{Rodriguez}, {Farr}, {Farr}, and
  {Mandel}}}]{rodriguez13}
\bibinfo{author}{\bibfnamefont{C.~L.} \bibnamefont{{Rodriguez}}},
  \bibinfo{author}{\bibfnamefont{B.}~\bibnamefont{{Farr}}},
  \bibinfo{author}{\bibfnamefont{W.~M.} \bibnamefont{{Farr}}},
  \bibnamefont{and} \bibinfo{author}{\bibfnamefont{I.}~\bibnamefont{{Mandel}}},
  \bibinfo{journal}{ArXiv e-prints}  (\bibinfo{year}{2013}{\natexlab{a}}),
  \eprint{arXiv:1308.1397}.

\bibitem[{\citenamefont{{Fairhurst}}(2009)}]{fairhurst2009}
\bibinfo{author}{\bibfnamefont{S.}~\bibnamefont{{Fairhurst}}},
  \bibinfo{journal}{New Journal of Physics} \textbf{\bibinfo{volume}{11}},
  \bibinfo{pages}{123006} (\bibinfo{year}{2009}), \eprint{arXiv:0908.2356}.

\bibitem[{\citenamefont{Wen and Chen}(2010)}]{wen2010}
\bibinfo{author}{\bibfnamefont{L.}~\bibnamefont{Wen}} \bibnamefont{and}
  \bibinfo{author}{\bibfnamefont{Y.}~\bibnamefont{Chen}},
  \bibinfo{journal}{Phys. Rev. D} \textbf{\bibinfo{volume}{81}},
  \bibinfo{pages}{082001} (\bibinfo{year}{2010}), \eprint{arXiv:1003.2504}.

\bibitem[{\citenamefont{{Fairhurst}}(2011)}]{fairhurst2011}
\bibinfo{author}{\bibfnamefont{S.}~\bibnamefont{{Fairhurst}}},
  \bibinfo{journal}{Class. Quantum Gravity} \textbf{\bibinfo{volume}{28}},
  \bibinfo{pages}{105021} (\bibinfo{year}{2011}), \eprint{arXiv:1010.6192}.

\bibitem[{\citenamefont{{Aasi, J et al. ( {LIGO Scientific Collaboration} and
  {Virgo Collaboration})}}(2013)}]{LVC2013}
\bibinfo{author}{\bibnamefont{{Aasi, J et al. ( {LIGO Scientific Collaboration}
  and {Virgo Collaboration})}}} (\bibinfo{year}{2013}),
  \eprint{arXiv:1304.0670}.

\bibitem[{\citenamefont{Gilks et~al.}(1996)\citenamefont{Gilks, Richardson, and
  Spiegelhalter}}]{gilks1996}
\bibinfo{author}{\bibfnamefont{W.}~\bibnamefont{Gilks}},
  \bibinfo{author}{\bibfnamefont{S.}~\bibnamefont{Richardson}},
  \bibnamefont{and}
  \bibinfo{author}{\bibfnamefont{D.}~\bibnamefont{Spiegelhalter}},
  \emph{\bibinfo{title}{\it{Markov Chain Monte Carlo in Practice}}},
  Interdisciplinary Statistics Series (\bibinfo{publisher}{Chapman and Hall},
  \bibinfo{year}{1996}), ISBN \bibinfo{isbn}{9780412055515}.

\bibitem[{\citenamefont{{Sidery, T. et. al.}}(2013)}]{sidery13}
\bibinfo{author}{\bibnamefont{{Sidery, T. et. al.}}}, \bibinfo{journal}{{In
  Preparation }}  (\bibinfo{year}{2013}).

\bibitem[{\citenamefont{{Rodriguez}
  et~al.}(2013{\natexlab{b}})\citenamefont{{Rodriguez}, {Farr}, {Raymond},
  {Farr}, {Littenberg}, {Fazi}, and {Kalogera}}}]{rodriguez13a}
\bibinfo{author}{\bibfnamefont{C.~L.} \bibnamefont{{Rodriguez}}},
  \bibinfo{author}{\bibfnamefont{B.}~\bibnamefont{{Farr}}},
  \bibinfo{author}{\bibfnamefont{V.}~\bibnamefont{{Raymond}}},
  \bibinfo{author}{\bibfnamefont{W.~M.} \bibnamefont{{Farr}}},
  \bibinfo{author}{\bibfnamefont{T.}~\bibnamefont{{Littenberg}}},
  \bibinfo{author}{\bibfnamefont{D.}~\bibnamefont{{Fazi}}}, \bibnamefont{and}
  \bibinfo{author}{\bibfnamefont{V.}~\bibnamefont{{Kalogera}}},
  \bibinfo{journal}{ArXiv e-prints}  (\bibinfo{year}{2013}{\natexlab{b}}),
  \eprint{arXiv:1309.3273}.

\bibitem[{\citenamefont{{Sidery} et~al.}(2013)\citenamefont{{Sidery}, {Gair},
  {Mandel}, and {Farr}}}]{sideryKD}
\bibinfo{author}{\bibfnamefont{T.}~\bibnamefont{{Sidery}}},
  \bibinfo{author}{\bibfnamefont{J.}~\bibnamefont{{Gair}}},
  \bibinfo{author}{\bibfnamefont{I.}~\bibnamefont{{Mandel}}}, \bibnamefont{and}
  \bibinfo{author}{\bibfnamefont{W.}~\bibnamefont{{Farr}}},
  \bibinfo{journal}{{In Preparation }}  (\bibinfo{year}{2013}).

\bibitem[{\citenamefont{Barack and Cutler}(2004)}]{barack2004}
\bibinfo{author}{\bibfnamefont{L.}~\bibnamefont{Barack}} \bibnamefont{and}
  \bibinfo{author}{\bibfnamefont{C.}~\bibnamefont{Cutler}},
  \bibinfo{journal}{Phys. Rev. D} \textbf{\bibinfo{volume}{69}},
  \bibinfo{pages}{082005} (\bibinfo{year}{2004}), \eprint{arXiv:gr-qc/0310125}.

\bibitem[{\citenamefont{{The LIGO Scientific Collaboration} and {The Virgo
  Collaboration}}(2012)}]{ligo12}
\bibinfo{author}{\bibnamefont{{The LIGO Scientific Collaboration}}}
  \bibnamefont{and} \bibinfo{author}{\bibnamefont{{The Virgo Collaboration}}},
  \bibinfo{journal}{ArXiv e-prints}  (\bibinfo{year}{2012}),
  \eprint{arXiv:1203.2674}.

\bibitem[{\citenamefont{{Buonanno} et~al.}(2009)\citenamefont{{Buonanno},
  {Iyer}, {Ochsner}, {Pan}, and {Sathyaprakash}}}]{buonanno09}
\bibinfo{author}{\bibfnamefont{A.}~\bibnamefont{{Buonanno}}},
  \bibinfo{author}{\bibfnamefont{B.~R.} \bibnamefont{{Iyer}}},
  \bibinfo{author}{\bibfnamefont{E.}~\bibnamefont{{Ochsner}}},
  \bibinfo{author}{\bibfnamefont{Y.}~\bibnamefont{{Pan}}}, \bibnamefont{and}
  \bibinfo{author}{\bibfnamefont{B.~S.} \bibnamefont{{Sathyaprakash}}},
  \bibinfo{journal}{Phys. Rev. D} \textbf{\bibinfo{volume}{80}},
  \bibinfo{pages}{084043} (\bibinfo{year}{2009}), \eprint{arXiv:0907.0700}.

\bibitem[{\citenamefont{{Fairhurst}}(2013)}]{fairhurst2013}
\bibinfo{author}{\bibfnamefont{S.}~\bibnamefont{{Fairhurst}}},
  \bibinfo{journal}{{In Preparation }}  (\bibinfo{year}{2013}).

\end{thebibliography}
\bibliographystyle{apsrev}
	
\end{document}